\documentclass[review]{elsarticle}

\usepackage{lineno,hyperref}
\modulolinenumbers[5]
\usepackage[utf8]{inputenc}
\usepackage{bm}
\usepackage{amssymb}
\usepackage{amsmath}
\usepackage{suffix}
\usepackage{amsfonts}
\usepackage{relsize}
\usepackage{mathtools}
\usepackage[makeroom]{cancel}
\usepackage{breqn}

\renewcommand{\text}[1]{{\rm #1}}
\newcommand{\smatrix}[2]{\left#2\begin{array}{#1}}
\newcommand{\ematrix}[1]{\end{array}\right#1}

\def\schrek{Schr\"odinger}
\def\rb{\bm{r}}
\def\nb{\bm{n}}
\def\ub{\bm{u}}
\def\fb{\bm{f}}
\def\Pb{\bm{P}}
\def\Bb{\bm{B}}
\def\Rb{\bm{R}}

\def\Cb{\bm{C}}

\def\Kb{\bm{K}}
\def\Mb{\bm{M}}
\def\Vb{\bm{V}}

\def\difd#1{\ {\rm d}#1}
\def\der#1#2{\frac{{\rm d}#1}{{\rm d}#2}}

\def\figdir{figures}

\newcommand{\tx}{\rm}
\newcommand{\uiL}{\ensuremath{^{\tx \scriptscriptstyle \stackrel{ion}{LOC}}}}
\newcommand{\uL}{\ensuremath{^{\tx \scriptscriptstyle LOC}}}
\newcommand{\uiN}{\ensuremath{^{\tx \scriptscriptstyle NL}}}

\renewcommand{\vec}[1]{\mathbf #1}
\newcommand{\dd}{{\rm d}}

\def\dr{\dd \vec{r}}

\newcommand\intr[2][\int]{#1 \! \dr \; #2}

\def\presuper#1#2%
  {\mathop{}%
   \mathopen{\vphantom{#2}}^{#1}%
   \kern-\scriptspace%
   #2}

\def\Vloc{
\uiL
}

\def\Vloc{\hat V\uL}

\def\Vloca{\hat V\uL _\alpha}

\def\Vnloc{\hat V\uiN}

\def\Vxc{V_{\tx xc}}

\def\Vexc{V_{\tx xc}^{\tx E}}

\def\rh{\rho}

\newcommand\deriv[2]{\frac{\partial}{\partial #2}\left(#1\right)}

\journal{Journal of Computational Physics}

\begin{document}

\begin{frontmatter}

\title{Finite element method and isogeometric analysis in electronic
  structure calculations: convergence study}

\author[NTC]{Robert Cimrman\corref{rc}}
\cortext[rc]{Corresponding author}
\ead{cimrman3@ntc.zcu.cz}
\author[KME,IP]{Matyáš Novák}
\author[IT]{Radek Kolman}
\author[ICS]{Miroslav Tůma}
\author[IP]{Jiří Vackář}

\address[NTC]{New Technologies Research Centre, University of West Bohemia,
  Univerzitní 8, 306 14 Plzeň, Czech Republic}

\address[KME]{Department of Mechanics, Faculty of Applied Sciences, University
  of West Bohemia, Univerzitní 22, 306 14 Plzeň, Czech Republic}

\address[IP]{Institute of Physics, Academy of Sciences of the Czech Republic,
  Na Slovance 1999/2, Prague, Czech Republic}

\address[IT]{Institute of Thermomechanics, Academy of Sciences of the Czech
  Republic, Dolejškova 5, 182 00 Prague, Czech Republic}

\address[ICS]{Institute of Computer Science, Academy of Sciences of the Czech
  Republic, Pod Vodárenskou věží 2, 182 07, Prague, Czech Republic}

\begin{abstract}
  We compare convergence of isogeometric analysis (IGA), a spline modification
  of finite element method (FEM), with FEM in the context of our real space
  code for ab-initio electronic structure calculations of non-periodic systems.
  The convergence is studied on simple sub-problems that appear within the
  density functional theory approximation to the \schrek{} equation: the
  Poisson problem and the generalized eigenvalue problem. We also outline the
  complete iterative algorithm seeking a fixed point of the charge density of a
  system of atoms or molecules, and study IGA/FEM convergence on a benchmark
  problem of nitrogen atom.
\end{abstract}


\end{frontmatter}

\linenumbers

\section{Introduction}

The electronic structure calculations are a rigorous tool for predicting and
understanding important properties of materials, such as elasticity, hardness,
electric and magnetic properties, etc. Those properties are tightly bound to
the notion of the total internal energy of a system of atoms --- a crucial
quantity to compute, and to determine its sensitivity w.r.t. various
parameters, e.g., the atomic positions in order to reach a stable arrangement.

Our team is developing a real space code \cite{ptcp} for electronic structure
calculations based on
\begin{itemize}
\item the density functional theory (DFT), \cite{DFT-1, DFT-2, DFT-3, DFT-4};
\item the environment-reflecting pseudopotentials \cite{vackar};
\item a weak solution of the Kohn-Sham equations \cite{Kohn-Sham}.
\end{itemize}
The code is based on the open source finite element package SfePy
\cite{SfePy-1} (Simple Finite Elements in Python, \url{http://sfepy.org}),
which is a general package for solving (systems of) partial differential
equations (PDEs) by the finite element method (FEM),
cf.~\cite{FEM-2}. Recently, it has been extended with the isogeometric analysis
(IGA)~\cite{IGA-1} is a spline-based modification of FEM. The key motivation
for this extension, besides interesting convergence properties~\cite{IGA-8} in
eigenvalue problems, was the possibility of a continuous field approximation
with a high global continuity on a simple domain --- a single NURBS
(Non-uniform Rational B-spline) patch. This feature is crucial for an efficient
evaluation of the sensitivity of the total energy w.r.t. a parameter, also
called the Hellman-Feynman forces (HFF)~\cite{IGA-FEM-Cimrman-1}.

Recently, using FEM and its variants in electronic structure calculation
context is pursued by a growing number of groups, cf.~\cite{DFT-FEM-Davydov},
where the $hp$-adaptivity is discussed, \cite{DFT-FEM-Motamarri-1,
  DFT-FEM-Motamarri-2} where spectral finite elements as well as the
$hp$-adaptivity are considered, or \cite{DFT-IGA-Masud}, where NURBS-based FEM
is applied.

IGA is a modification of FEM which employs shape functions of different spline
types such as B-splines, NURBS \cite{NURBS}), T-splines \cite{IGA-6}, etc. It
was successfully employed for numerical solutions of various physical and
mathematical problems, such as fluid dynamics, diffusion and other problems of
continuum mechanics~\cite{IGA-1, IGA-4, IGA-8}. The theoretical works relating
to the convergence behaviour of IGA have been published in~\cite{IGA-9, IGA-10,
  IGA-7, IGA-11}.

The drawbacks of using IGA, as reported in \cite{IGA-8}, concern mainly the
increased computational cost of the numerical integration and assembling. Also,
because of the higher global continuity, the assembled matrices have more
nonzero entries than the matrices corresponding to the $C^0$ FEM basis. A
comparison study of IGA and FEM matrix structures, the cost of their
evaluation, and mainly the cost of direct and iterative solvers in IGA has been
presented by \cite{IGA-Collier} and \cite{IGA-Schillinger-1}.

In this paper we compare numerical convergence properties of FEM and IGA using
problems originating from various stages of our electronic structure
calculation algorithm, in order to assess the applicability of IGA for our
purposes. It is structured as follows: in Section~\ref{sec:esc} we provide a
light-weight introduction to the topic of electronic structure calculations, in
Section~\ref{sec:dm} the used discretization methods (FEM and IGA) are
presented. Finally, in Section~\ref{sec:results} the numerical convergence
results are presented: first for several Poisson problems because the Poisson
problem solution is an important part of the algorithm used for calculation of
the electrostatic potential (see below); then several eigenvalue problems
corresponding to simple quantum mechanical systems with a similar matrix
structure to that of the complete problem; and finally, for the overall
algorithm.

\section{Electronic structure calculations}
\label{sec:esc}

Let us briefly introduce the topic of electronic structure calculations. The
systems of atoms and molecules are described in the most general form by the
many-particle \schrek{} equation, cf.~\cite{DFT-3},
\begin{equation}
  \label{eq:many-part}
  H \Psi(e_1, e_2, \dots, e_n) = \varepsilon \Psi(e_1, e_2, \dots, e_n) \;,
\end{equation}
where $H$ is the Hamiltonian (energy operator) of the system, $e_i$ the
particles (e.g. electrons) and $\varepsilon$ the energy of the state
$\Psi$. The equation (\ref{eq:many-part}) is, however, too complicated to
solve, even for three electrons. Among the techniques reducing this complexity,
we use the DFT approach \cite{DFT-1}. The DFT allows decomposing the
many-particle \schrek{} equation into the one-electron Kohn-Sham equations
\cite{Kohn-Sham}. Using atomic units they can be written in the common form
\begin{equation}
  \label{eq:kohn-sham}
  \left(-\frac{1}{2} \nabla^2 + V_{\text{H}}(\rb) + V_{\text{xc}}(\rb)
    + \hat{V}(\rb) \right) \psi_i = \varepsilon_i \psi_i \;,
\end{equation}
which provide the orbitals $\psi_i$ that reproduce, with the weights of
occupations $n_i$, the charge density $\rho$ of the original interacting
system, as
\begin{equation}
  \label{eq:density}
  \rho(\rb) = \sum_i^N n_i |\psi_i(\rb)|^2 \;.
\end{equation}
$\hat{V}$ is a (generally) non-local Hermitian operator representing the
effective ionic potential for electrons. In the present case, within
pseudopotential approach, $\hat{V}$ represents core electrons, separated from
valence electrons, together with the nuclear charge. $V_{\text{xc}}$ is the
exchange-correlation potential describing the non-coulomb electron-electron
interactions. The exact potential is not known, so we use local-density
aproximation (LDA) of this potential \cite{DFT-3}, where the potential is
a function of charge density at a given point. $V_{\text{H}}$ is the
electrostatic potential obtained as a solution to the Poisson equation. The
Poisson equation for $V_{\text{H}}$ has the charge density $\rho$ at its
right-hand side and is as follows:
\begin{equation}
  \label{eq:poisson}
  \Delta V_{\text{H}} = 4 \pi \rho \;.
\end{equation}
Denoting the total potential $V := V_{\text{H}} + V_{\text{xc}} + \hat{V}$, we
can write, using Hartree atomic units,
\begin{equation}
  \label{eq:kohn-sham2}
  \left(-\frac{1}{2} \nabla^2 + V(\rb) \right) \psi_i = \varepsilon_i \psi_i \;.
\end{equation}
Note that the above mentioned eigenvalue problem is highly non-linear, as the
potential $V$ depends on the orbitals $\psi_i$. Therefore an iterative scheme
is needed, defining the DFT loop for attaining a self-consistent solution.

\subsection{DFT loop}

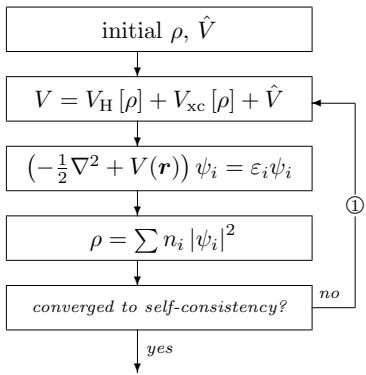
\begin{figure}[h]
  \vspace*{20pt}
  \begin{center}
    \setlength{\unitlength}{0.85ex}
    \scalebox{0.9}{
      \begin{picture}(60,50)
        \put(10,50){\framebox(35,5){initial $\rho$, $\hat{V}$}}
        \put(25,50){\vector(0,-1){3}}
        \put(10,42){\framebox(35,5){$V = V_{\text{H}} \left [ \rho \right ]
            + V_{\text{xc}} \left [ \rho \right ] + \hat{V}$}}
        \put(25,42){\vector(0,-1){3}}
        \put(10,34){\framebox(35,5){$\left(-\frac{1}{2} \nabla^2
              + V(\rb) \right) \psi_i = \varepsilon_i \psi_i$}}
        \put(25,34){\vector(0,-1){3}}
        \put(10,26){\framebox(35,5){
            $\rho = \sum n_i \left | \psi_i \right |^2$}}
        \put(25,26){\vector(0,-1){3}}
        \put(10,18){\framebox(35,5){\scriptsize \it converged to
            self-consistency?}}

        \put(45,20.5){\line(1,0){5}}
        \put(50,20.5){\line(0,1){10.5}}
        \put(50,32.25){\circle{2}}
        \put(50,32.25){\makebox(0,0){\scriptsize 1}}
        \put(50,33.5){\line(0,1){10.5}}
        \put(50,44.0){\vector(-1,0){5}}

        \put(45,20.5){\makebox(4,3){\scriptsize \it no}}
        \put(25,18){\vector(0,-1){5}}
        \put(25,14.0){\makebox(5,3){\scriptsize \it yes}}
      \end{picture}
    }
  \end{center}
  \vspace*{-20mm}
  \caption{DFT, iterative self-consistent scheme.} \label{fig:scheme}
\end{figure}

For the global convergence of the DFT iteration we use the standard algorithm
outlined in Fig.~\ref{fig:scheme}. The purpose of the DFT loop is to find a
self-consistent solution --- a fixed point of a function of the charge density
$\rho$. For this task, a variety of nonlinear solvers can be used. We use
Broyden-type quasi-Newton solvers applied to
\begin{equation}
  \label{eq:dft}
  DFT(\rho^{i}) - \rho^{i} = \rho^{i+1} - \rho^{i} = 0 \;,
\end{equation}
where $DFT$ denotes a single iteration of the DFT loop.

After the DFT loop convergence is achieved, the derived quantities,
particularly the total energy, are computed. By minimizing the total energy as
a function of atomic positions, the equilibrium atomic positions can be found.
Therefore the DFT loop itself can be embedded into an outer optimization loop,
where the objective function gradients are the HFF.

\subsection{Total Energy and Forces Acting on Atoms}

The total energy of the system can be obtained within DFT as the sum of the
ion-ion interaction energy (i.e. energy of electrostatic interations among
nuclei), the kinetic energy of electrons, the electron-ion interaction energy,
the electron-electron electrostatic interaction energy and the exchange and
correlation energy that reflect the fact that electron, as any fermions,
satisfy the Pauli principle and the fact that electrons, like any
quantum-mechanical particles, are indistinguishable in principle (i.e. via an
exchange of two electrons we don't get another quantum state). The terms
mentioned above, respectively, can be expressed as
\begin{eqnarray}
  E_{\tx tot} =
  \frac{1}{2} \sum_{a,a^\prime\neq a}
  \frac{Z_\alpha Z_{\alpha^\prime}}{\left| \tau_\alpha
      - \tau_{\alpha^\prime}\right|}
  +
  \sum_i n_i \intr{ \; \psi^*_i(\vec{r})
    \left(-\frac{1}{2}\nabla^2\right)\psi_i(\vec{r})}
  + \nonumber \\
  +
  \intr{ \rh(\vec{r}) \hat V(\vec{r}) }
  +
  \frac{1}{2} \int\!\!\! \intr{
    \frac{\rh(\vec{r}) \rh(\vec{r^\prime})}
    {\left|\vec{r}-\vec{r^\prime}\right|} }
  +
  \intr{  \rh(\vec{r}) \Vexc(\vec{r};\rh)} \; ,
\end{eqnarray}
where $\alpha$ refers to atomic sites and $Z$ stands for the ionic charge of
the nucleus (or of the core, in case of pseudopotentials). $\Vexc(\vec{r};\rh)$
denotes the exchange-correlation energy functional of the charge density
related to the exchange-correlation potential via
\begin{equation}
\Vxc =  \deriv{\Vexc}{\rho} \; .
\end{equation}

The force acting on atom $\alpha$ is equal to the derivative of the total
energy functional with respect to an infinitesimal displacement of this atom
$\delta\tau_\alpha$:
\begin{equation}
  \label{eq:force1}
  {\mathbf F}^\alpha = - \frac{\delta E}{\delta\tau_\alpha}
\end{equation}

Making use of the Hellmann-Feynman theorem that relates the derivative of the
total energy with respect to a parameter $\lambda$, to the expectation value of
the derivative of the Hamiltonian operator w.r.t. the same parameter
\begin{equation}
  \label{eq:hft}
  \der{E}{\lambda}
  = \left\langle \psi^{*}_{\lambda}
  \left| \der{\hat{H}_\lambda}{\lambda} \right|
  \psi_\lambda \right\rangle \rm ,
\end{equation}
within the density functional theory we can write
\begin{equation}
  \label{eq:force2}
  {\mathbf F}^\alpha = {\mathbf F}^\alpha_{\rm HF,es} -
  \frac{\left( \sum_i n_i \delta\varepsilon_i -
      \int \rho({\mathbf r}) \delta \left[\hat V + V_{\rm H}
        + V_{\rm xc}\right]\! ({\mathbf r}) \dd^3 {\mathbf r}
    \right)}{\delta \tau_\alpha}
\end{equation}
where the first term is the electrostatic Hellmann-Feynman force (formed by the
sum over all the atoms $\beta\neq\alpha$ of electrostatic forces between the
charges of atomic nuclei $Z_\alpha$ and $Z_\beta$ and by the force acting on
the charge $Z_\alpha$ in the charge density $\rho$)
\begin{equation}
  \label{eq:hff-es}
  {\mathbf F}^\alpha_{\rm HF,es} = Z_\alpha \der{}{\tau_\alpha}
  \left(
    - \sum_{\beta\neq\alpha} \frac{Z_\beta}{|\tau_\alpha - \tau_\beta|}
    + \int \frac{\rho({\mathbf r})}{|\tau_\alpha-{\mathbf r}|} d^3 {\mathbf r}
  \right)
\end{equation}
and the second term in Eq.(\ref{eq:force2}), is the ``Pulay'' force, also known
as ``incomplete basis set'' force, that contains the corrections that depend on
technical details of the calculation and can be extremely complicated to
evaluate for some non-tirivial types of bases. By means of the fixed basis
independent of atomic position and via the wave function $\psi_i$ continuity up
to second derivatives, we can get rid of this troublesome term. Even without
this term, the evaluation of pure HF-force in case of non-local separable
pseudopotentials acting in the $l-$projected (via the spheriacal harmonics)
subspaces,
\begin{eqnarray}
{\mathbf F}^\alpha_{\rm HF,es} =
\deriv{\intr{ \rh \Vloc }}{\tau_\alpha} +
   \deriv{\sum_i n_i \int \dd\vec{r} \; \psi^*_i \Vnloc \psi_i}{\tau_\alpha}
   \nonumber\\
   + \deriv{   \frac{1}{2} \int \dd\vec{r} \; \rho\uiL \Vloc}{\tau_\alpha}
   - \deriv{   \frac{1}{2} \sum_a \int\dd\vec{r} \;
     \rho\uiL_a \Vloca }{\tau_\alpha} \; ,
\end{eqnarray}
where the superscripts $\ensuremath{^{\tx \scriptscriptstyle LOC}}$ and
$\ensuremath{^{\tx \scriptscriptstyle NL}}$ denote the local and non-local
pseudopotential parts, respectively, and
\begin{equation}
  \rho\uiL(\vec{x}) = -\frac{1}{4\pi} \nabla^2 \Vloc(\vec{x}) \; ,
\end{equation}
might be non-trivial, as it was shown by Ihm, Zunger and
Cohen\cite{IhmZungerCohen} (for more details see e.g. \cite{Krakauer},
\cite{Weinert}, \cite{Alessandra}). But, anyway, it seems to be much more
acceptable for practical use than the evaluating the additional Pulay term
within finite-element basis would be.


\section{Discretization methods}
\label{sec:dm}

Before presenting key points of FEM and IGA, our problem needs to be reinstated
in a weak form, usual in the finite element setting.

\subsection{Weak formulation}
\label{sec:weak}

Let us denote $H^1(\Omega)$ the usual Sobolev space of functions with $L^2$
integrable derivatives and $H^1_0(\Omega) = \{u \in H^1(\Omega)
| u = 0 \mbox{ on } \partial \Omega\}$.

The eigenvalue problem (\ref{eq:kohn-sham2}) can be rewritten using the weak
formulation: find functions $\psi_i \in H^1(\Omega)$ such that for all $v \in
H^1_0(\Omega)$ holds
\begin{equation}
  \label{eq:kohn-sham-weak}
  \int_\Omega \frac{1}{2} \nabla \psi_i \cdot \nabla v \difd{V}
  + \int_\Omega v V \psi_i \difd{V}
  = \varepsilon_i \int_\Omega v \psi_i \difd{V}
  + \oint_{\partial \Omega} \frac{1}{2} \der{\psi_i}{\nb} \difd{S} \;.
\end{equation}
If the solution domain $\Omega$ is sufficiently large, the last term
can be neglected. The Poisson equation (\ref{eq:poisson}) has the following
weak form:
\begin{equation}
  \label{eq:poisson-weak}
  \int_{\Omega} \nabla v \cdot \nabla V_{\text{H}} = 4 \pi \int_{\Omega} \rho v
  \;.
\end{equation}

Equations (\ref{eq:kohn-sham-weak}), (\ref{eq:poisson-weak}) then need to be
discretized --- the continuous fields are approximated by discrete fields with
a finite set of degrees of freedom (DOFs) and a basis, typically piece-wise
polynomial:
\begin{equation}
  \label{eq:discretize}
  u(\rb) \approx u^h(\rb) = \sum_{k=1}^{N} u_k \phi_k(\rb)
  \mbox{ for } \rb \in \Omega \;,
\end{equation}
where $u$ is a continuous field ($\psi$, $v$, $V_H$ in our equations), $u_k$,
$k = 1, 2, \dots, N$ are the discrete DOFs and $\phi_k$ are the basis
functions. From the computational point of view it is desirable that the basis
functions have a small support, so that the resulting system matrix is sparse.

Substituting (\ref{eq:discretize}) into (\ref{eq:kohn-sham-weak}) leads to the
matrix form of the Kohn-Sham eigenvalue problem:
\begin{equation}
  \label{eq:kohn-sham-discrete}
  \left(\Kb + \Vb(\psi_i)\right) \psi_i = \varepsilon_i \Mb \psi_i \;,
\end{equation}
where
\begin{eqnarray*}
  && \Kb = \{K_{ij}\} \equiv \int_{\Omega_h} \nabla \phi_i \nabla \phi_j \;, \\
  && \Vb(\psi_i) = \{V_{ij}\} \equiv \int_{\Omega_h} \phi_i V(\psi_i) \phi_j
  \;, \\
  && \Mb = \{M_{ij}\} \equiv \int_{\Omega_h} \phi_i \phi_j \;.
\end{eqnarray*}
Similarly,  the matrix form of the Poisson problem (\ref{eq:poisson-weak}) is:
\begin{equation}
  \label{eq:poisson-discrete}
  \Kb \ub = \fb \;,
\end{equation}
where $V_{\text{H}}(\rb) \approx \sum_{k=1}^{N} u_k \phi_k(\rb)$ and $\fb =
\{f_{i}\} \equiv 4 \pi \int_{\Omega_h} \rho \phi_i$.

\subsection{Finite element method}
\label{sec:fem}

In the FEM the discretization process involves the discretization of the domain
$\Omega$ --- it is replaced by a polygonal domain $\Omega_h$ that is covered by
small non-overlapping subdomains called \emph{elements} (e.g. triangles or
quadrilaterals in 2D, tetrahedrons or hexahedrons in 3D), cf.~\cite{FEM-1,
  FEM-2}. The elements form a finite element \emph{mesh}.

The basis functions are defined as piece-wise polynomials over the individual
elements, have a small support and are typically globally $C^0$ continuous. The
discretized equations are evaluated over the elements as well to obtain local
matrices or vectors that are then assembled into a global sparse system. The
evaluation usually involves a numerical integration on a reference element, and
a mapping to individual physical elements \cite{FEM-1, FEM-2}. The nodal basis
of Lagrange interpolation polynomials or the hierarchical basis of Lobatto
polynomials can be used in our code.

\subsection{Isogeometric analysis}
\label{sec:iga}

The basis functions in IGA are formed directly from the CAD geometrical
description in terms of NURBS patches, without the intermediate FE mesh --- the
meshing step is removed, which is one of its principal advantages. A NURBS
patch is a single NURBS object --- a linear combination of \emph{control
  points} $\Pb = \{\Pb_A\}_{A=1}^{N}$ (or unknown field coefficients) and NURBS
basis functions $R_{A,p}(\underline{\xi})$, where $p$ is the NURBS solid degree
and $\underline{\xi} = \{\xi_1, \dots, \xi_D\}$ are the parametric
coordinates. Thus, a d-dimensional geometric domain is defined by
\begin{equation}
  \label{eq:iga-1}
  \rb(\underline{\xi})
  = \sum_{A=1}^{n} \Pb_A R_{A,p}(\underline{\xi})
  = \Pb^T \Rb(\underline{\xi}) \;.
\end{equation}
If $d > 1$, the NURBS solid can be defined as a tensor product of univariate
NURBS curves. The basic properties of the B-spline basis functions can be
found in \cite{NURBS}.

The same NURBS basis is used also for the approximation of a continuous
field $u$ ($\psi$, $v$, $V_H$ in our equations):
\begin{equation}
  \label{eq:iga-1}
  \ub(\underline{\xi})
  = \sum_{A=1}^{n} u_A R_{A,p}(\underline{\xi}) \;,
\end{equation}
where $u_A$ are the unknown DOFs --- coefficients of the basis in the linear
combination.

Our implementation \cite{SfePy-2} uses a variant of IGA based on \emph{Bézier
  extraction operators} \cite{IGA-2} that is suitable for inclusion into
existing FE codes. The code itself does not see the NURBS description at
all. It is based on the observation that repeating a knot in the knot vector
decreases continuity of the basis in that knot by one. This can be done in such
a way that the overall shape remains the same, but the ``elements'' appear
naturally as given by non-zero knot spans. The final basis restricted to each
of the elements is formed by the Bernstein polynomials $\Bb$. The assembling of
matrices and vectors on resulting from (\ref{eq:kohn-sham-discrete}) then
proceeds in the usual FE sense (cf.~\cite{IGA-2}):
\begin{enumerate}
\item Setup points $\xi_q$ for numerical quadrature on a reference element.
\item Loop over elements of the Bézier ``mesh'' (given by knot spans).
\item On each element $e$:
  \begin{enumerate}
  \item Evaluate the Bernstein basis $\Bb(\xi_q)|_e$,
  \item Reconstruct the original NURBS basis: $\Rb(\xi_q)|_e = \Cb|_e
    \Bb(\xi_q)|_e$, using the Bézier extraction operator $\Cb$, that is local
    to element $e$.
  \item Evaluate element contributions to the global matrix.
  \item Assemble using the original DOF connectivity.
  \end{enumerate}
\end{enumerate}

The Bézier extraction matrices $\Cb$ are pre-computed for a NURBS patch domain
using an efficient algorithm that employs the tensor-product nature of the
patch \cite{IGA-2}, and then reused in all subsequent computations on that
domain.

\section{Results}
\label{sec:results}

The electronic structure calculations described in Section \ref{sec:esc}
involve solving the following two sub-problems:
\begin{itemize}
\item the Poisson equation (\ref{eq:poisson-weak}) for the potential $V_H$,
\item and the generalized eigenvalue problem (\ref{eq:kohn-sham-weak}).
\end{itemize}
Below we compare the convergence of FEM and IGA when applied to the two
sub-problems, as well as to the entire algorithm of the DFT loop
(\ref{eq:dft}).

All computations were done on a tensor-product domain, with varying number of
vertices/knots along an edge.

We are interested in convergence w.r.t. three parameters:
\begin{itemize}
\item The number of vertices/knots along the domain edge $N_e$ provides insight
  into the work necessary to integrate over the domain, because $(N_e - 1)^d$
  is equal to the number of elements/Bézier elements. Also, the element size
  $h$ can be obtained as $h = 1 / ((N_e - 1)$.
\item The number of non-zero entries in the matrices $N_{nz}$ that corresponds
  to the cost of matrix-vector products, and hence the cost of a single
  linear/eigenvalue problem solver iteration.
\item The total number of degrees of freedom $N_{dof}$, i.e. the size of the
  matrices, which is related to the difficulty of solving the Poisson equation
  (\ref{eq:poisson-weak}) or the eigenvalue problem (\ref{eq:kohn-sham-weak}).
\end{itemize}
Thus a higher $N_e$ indicates a higher cost of assembling the matrices, while
higher $N_{nz}$ and $N_{dof}$ mean a more difficult problem solution.

\subsection{Poisson's equation with manufactured solutions}
\label{sec:poisson}

In the method of manufactured solutions, cf.~\cite{IGA-11}, a solution $u(\rb)$
is made up and the left-hand side operator (the Laplace operator here) is
applied to obtain the corresponding right-hand side $g(\rb) \equiv \Delta
u(\rb)$. The function $g(\rb)$ is then used as the right-hand side in the
numerical solution, while $u(\rb)$ for $\rb \in \Gamma$ is applied as the
Dirichlet boundary condition on the whole domain surface $\Gamma$.

Several analytic formulas were considered in 1D, 2D and 3D. The domain was a
unit cube (or square): $\Omega \equiv [-0.5, 0.5]^d$, $d = 1, 2, 3$. The
discretization parameters are summarized in Tab.~\ref{tab:poisson-pars}.

In the FEM setting, the Lagrange basis with 1D polynomial orders 1, 2 and 3 and
the uniform discretization were used. In the IGA context, the B-spline basis
with 1D degrees 1, 2 and 3, and the approximation with uniformly distributed
knots, were used. We also varied the global continuity of the B-spline basis:
up to $C^2$, depending on the B-spline degree. Note that the B-spline basis
with $C^0$ continuity is equivalent to the FEM basis.

\begin{table}[ht!]
  \centering
  \begin{tabular}{r|lc}
    dimension & $N_e$ & solver \\
    \hline
    1D & 100, 200, 500, 1000, 2000 & direct \\
    2D & 5, 10, 15, 25, 40, 60 & iterative \\
    3D & 5, 10, 15, 20, 25 & iterative
  \end{tabular}
  \caption{Poisson problem: discretization parameters --- numbers of
    vertices/knots along the domain edge, and linear solver kind.}
  \label{tab:poisson-pars}
\end{table}

\begin{figure}[ht!]
  \centering

  \includegraphics[width=0.48\linewidth]{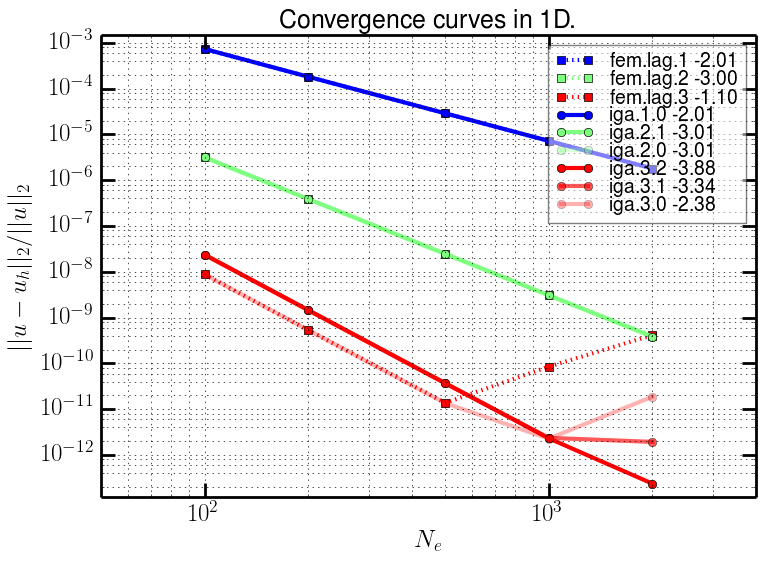}
  \includegraphics[width=0.48\linewidth]{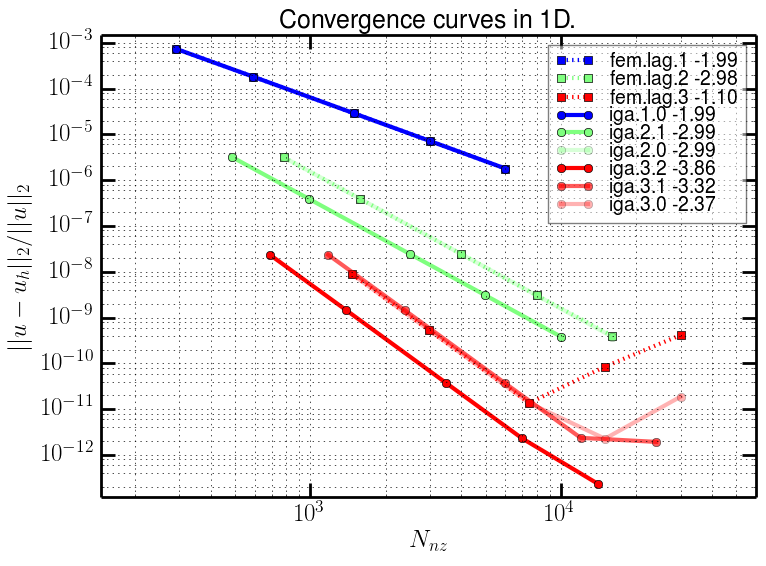}
  \includegraphics[width=0.48\linewidth]{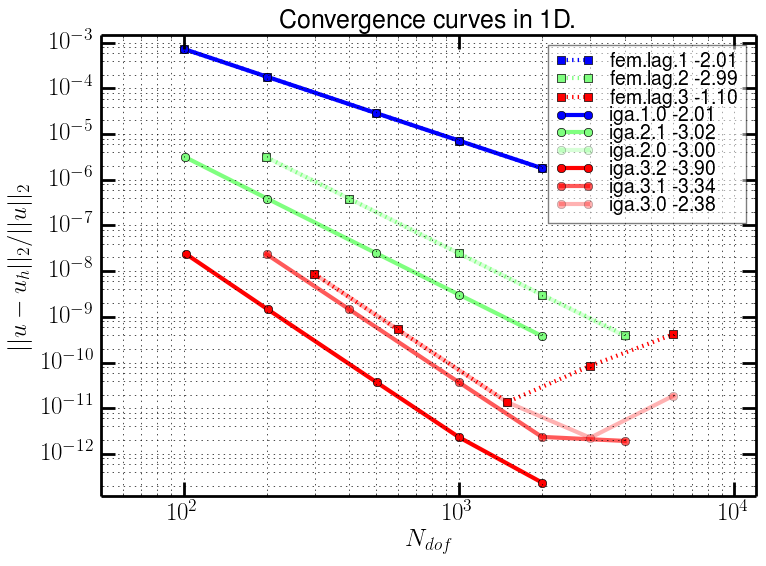}
  \includegraphics[width=0.48\linewidth]{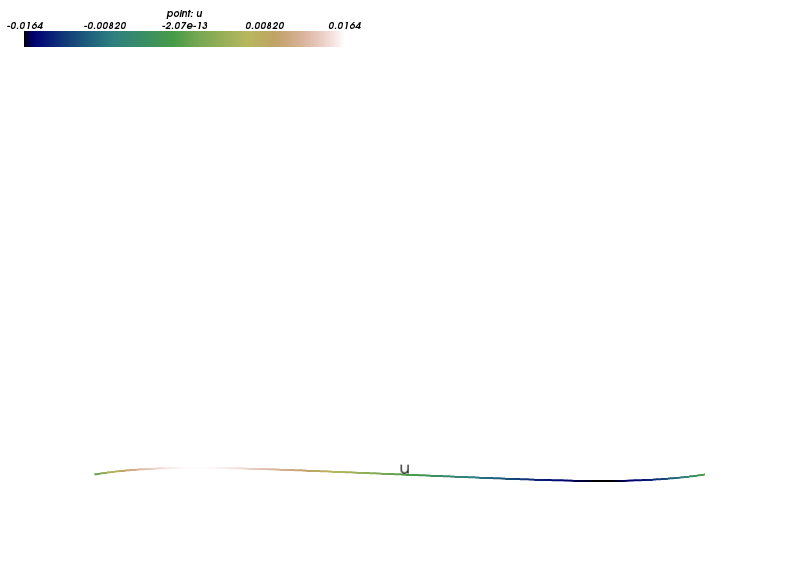}

  \caption{Convergence results for
    $u(x) \equiv \left(x^{4} - 0.0625\right) \sin{\left (x \right )}$.
    The legend for the FEM shows the polynomial order and the slope of the
    corresponding curve. For the IGA it shows the degree, the global
    continuity, and the slope, respectively. The function $u$ is shown in
    bottom-right.}
  \label{fig:1d-p04sin}
\end{figure}

\begin{figure}[ht!]
  \centering

  \includegraphics[width=0.48\linewidth]{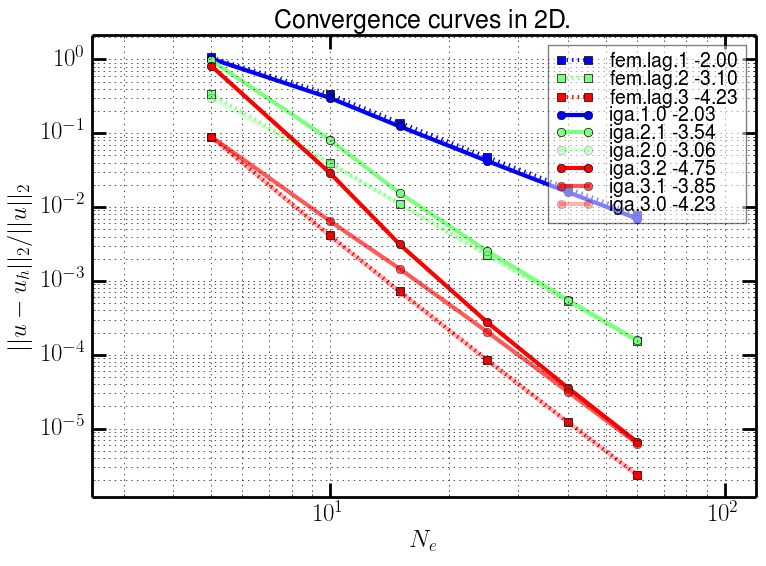}
  \includegraphics[width=0.48\linewidth]{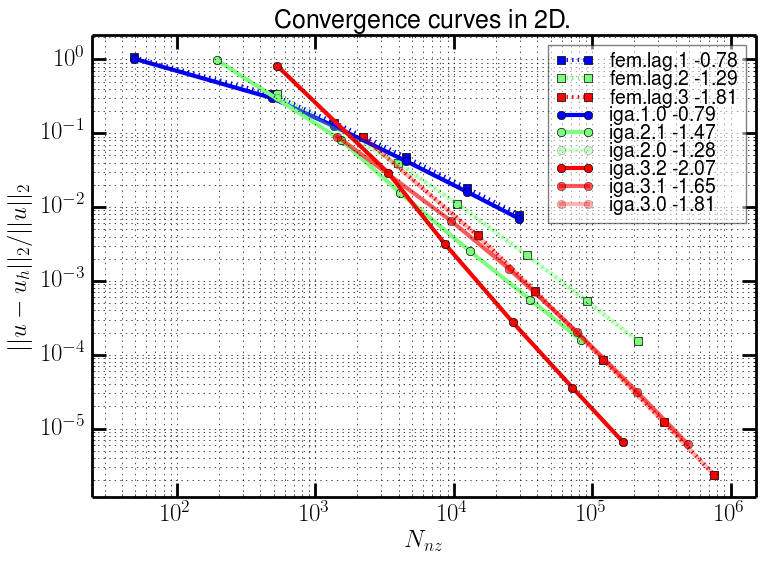}
  \includegraphics[width=0.48\linewidth]{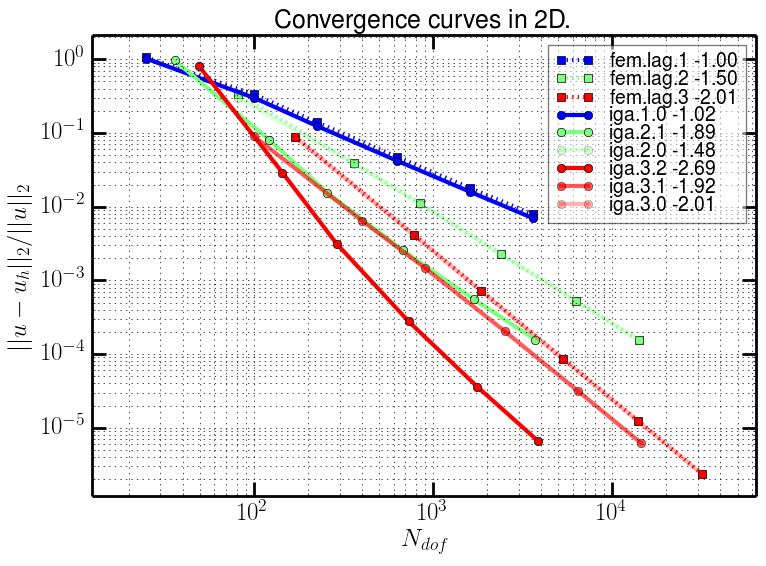}
  \includegraphics[width=0.48\linewidth]{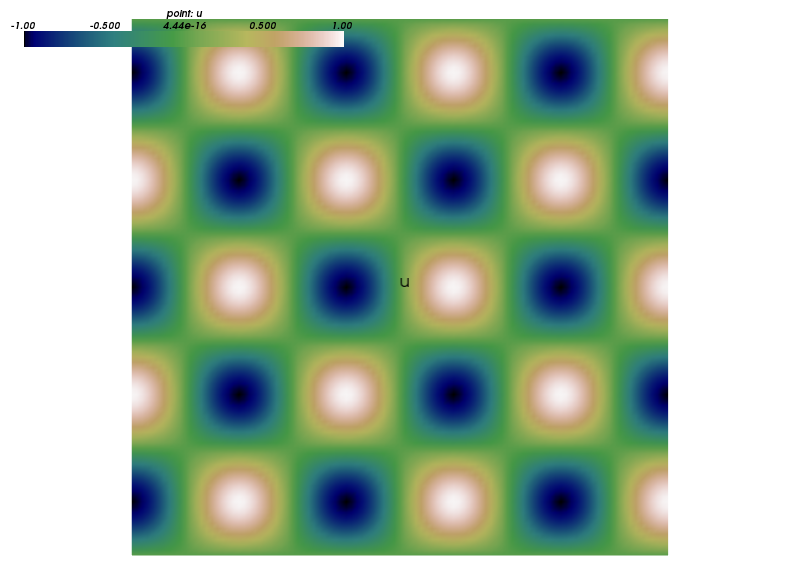}

  \caption{Convergence results for
    $u(x, y) \equiv \sin{\left (5 \pi x \right )} \cos{\left (5 \pi y \right )}$.
    The legend for the FEM shows the polynomial order and the slope of the
    corresponding curve. For the IGA it shows the degree, the global
    continuity, and the slope, respectively. The function $u$ is shown in
    bottom-right.}
  \label{fig:2d-sincos}
\end{figure}

\begin{figure}[ht!]
  \centering

  \includegraphics[width=0.48\linewidth]{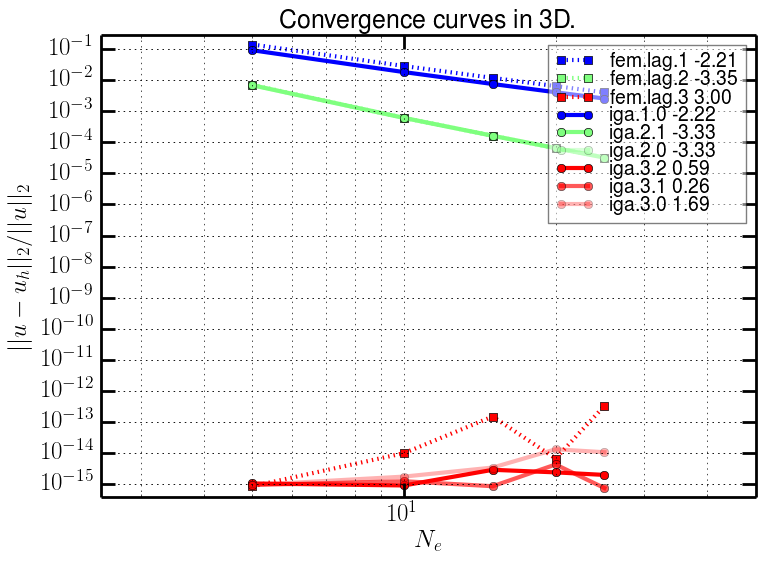}
  \includegraphics[width=0.48\linewidth]{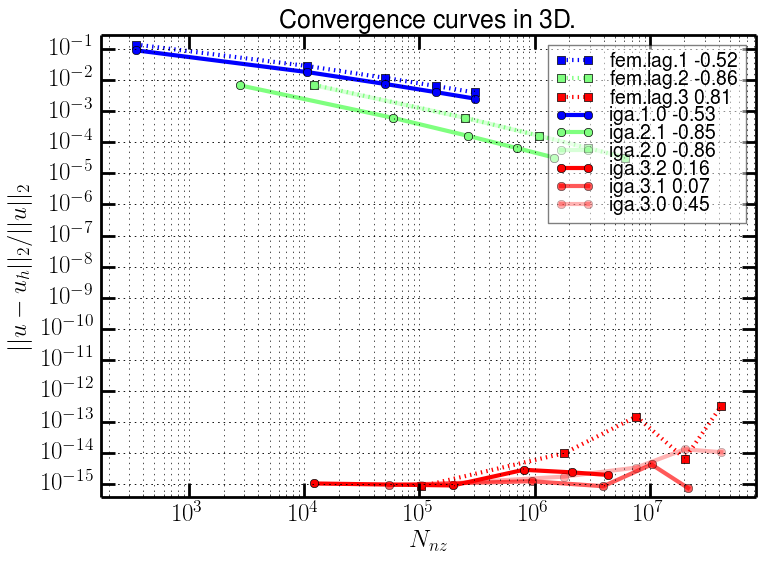}
  \includegraphics[width=0.48\linewidth]{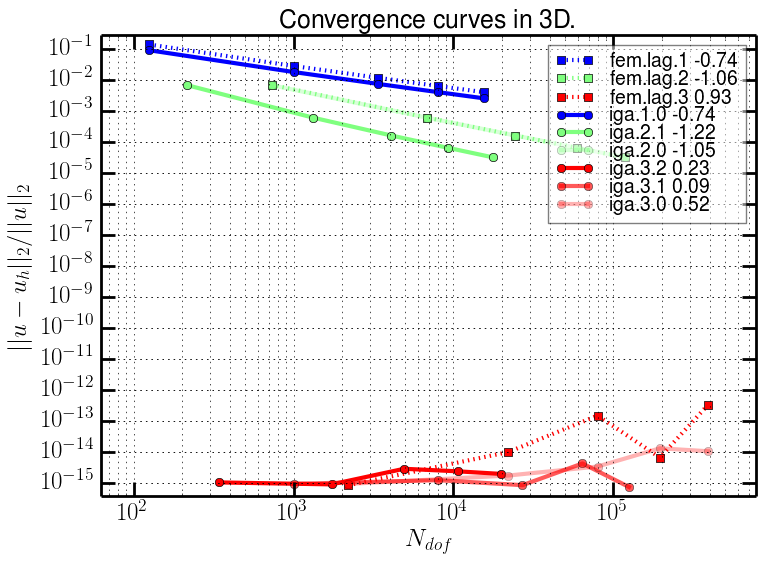}
  \includegraphics[width=0.48\linewidth]{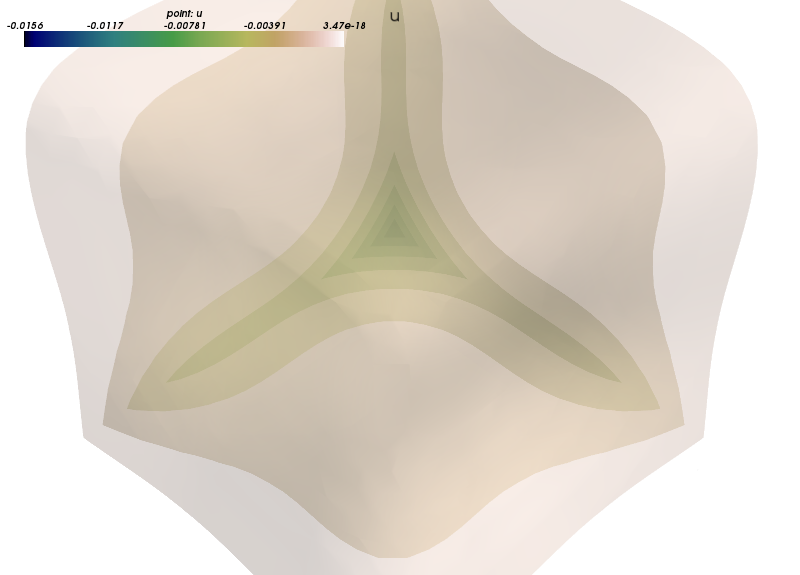}

  \caption{Convergence results for
    $u(x, y, z) \equiv \left(x^{3} - 0.125\right) \left(y^{3} - 0.125\right) \left(z^{3} - 0.125\right)$.
    The legend for the FEM shows the polynomial order and the slope of the
    corresponding curve. For the IGA it shows the degree, the global
    continuity, and the slope, respectively. The function $u$ is shown in
    bottom-right.}
  \label{fig:3d-p03}
\end{figure}

\begin{figure}[ht!]
  \centering

  \includegraphics[width=0.48\linewidth]{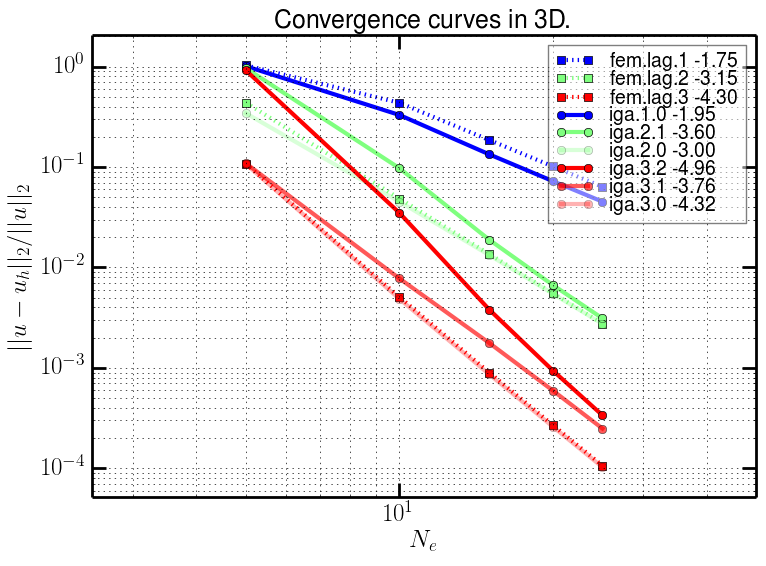}
  \includegraphics[width=0.48\linewidth]{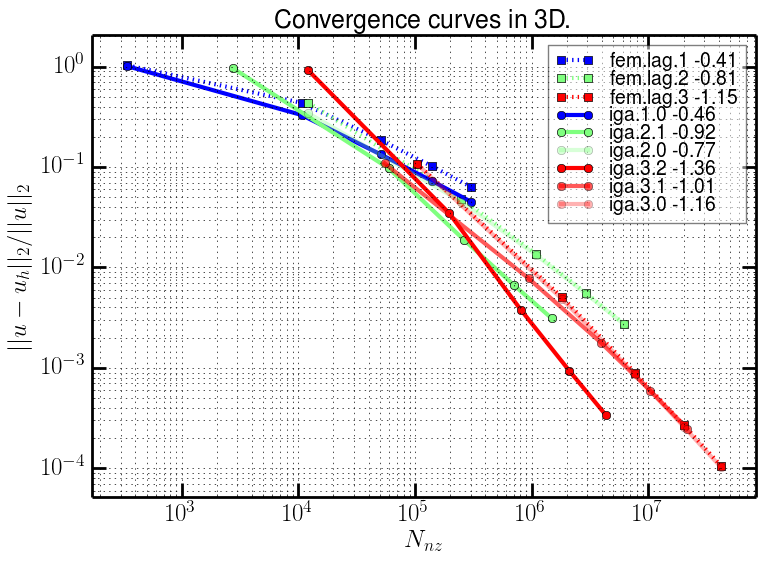}
  \includegraphics[width=0.48\linewidth]{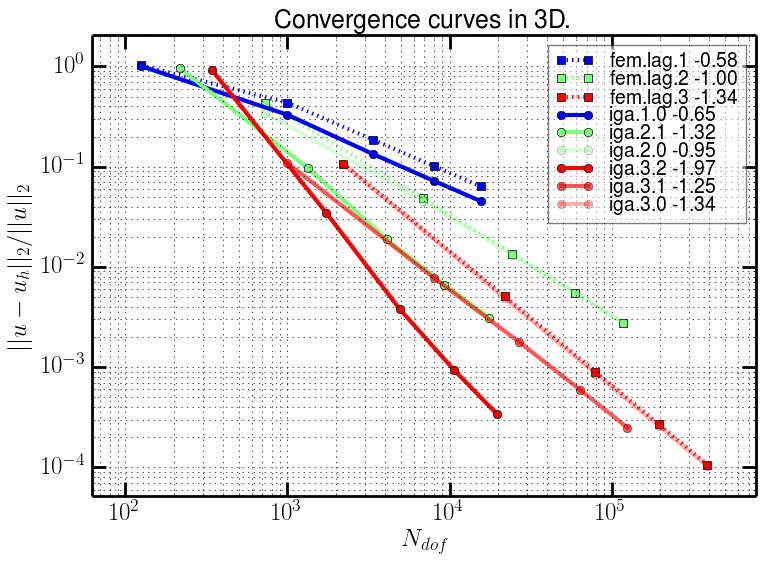}
  \includegraphics[width=0.48\linewidth]{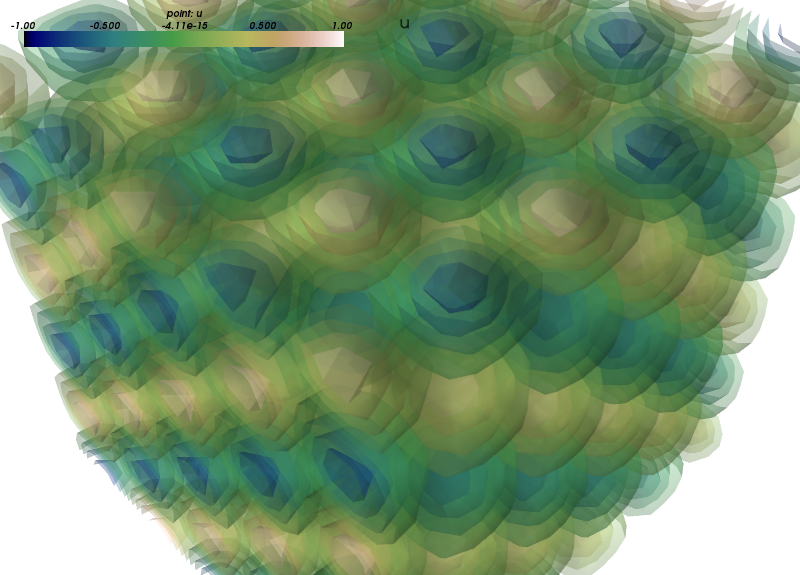}

  \caption{Convergence results for
    $u(x, y, z) \equiv \sin{\left (5 \pi x \right )} \sin{\left (5 \pi z \right )} \cos{\left (5 \pi y \right )}$.
    The legend for the FEM shows the polynomial order and the slope of the
    corresponding curve. For the IGA it shows the degree, the global
    continuity, and the slope, respectively. The function $u$ is shown in
    bottom-right.}
  \label{fig:3d-sincossin}
\end{figure}

The convergence of FEM and IGA for the Poisson problem is compared in
Figs.~\ref{fig:1d-p04sin}-~\ref{fig:3d-sincossin}. All the figures contain the
convergence curves w.r.t. $N_e$ (top left), $N_{nz}$ (top right) and $N_{dof}$
(bottom left). The relative error $||u - u_h||_2/||u||_2$ is measured. The
analytic solution $u$ is depicted in bottom right. The figure legends use the
following naming scheme:
\begin{itemize}
\item FEM Lagrange basis: ``fem.lag.$<$degree$>$ $<$slope$>$'';
\item IGA basis: ``iga.$<$degree$>$.$<$continuity$>$ $<$slope$>$''.
\end{itemize}

\subsubsection{1D Poisson problem}

In the 1D case (Fig.~\ref{fig:1d-p04sin}), $u(x) \equiv \left(x^{4} -
  0.0625\right) \sin{\left (x \right )}$ was used. Here, IGA with increasing
continuity performs progressively better then FEM with the same polynomial
order when the convergence w.r.t. $N_{nz}$ and $N_{dof}$ is considered. On the
other hand, its error is slightly higher than the FEM or $C^0$ IGA error for
the $N_e$ curve. The right parts of the curves for order/degree three solutions
exhibit a loss numerical precision for very small errors --- this loss of
precision seems to decrease considerably with increasing the IGA basis
continuity, and for the highest system resolution, IGA outperforms FEM even
w.r.t. $N_e$. This is probably related to a better conditioning of the
resulting linear system. Note that a direct solver \cite{UMFPACK} (version
5.6.2) was used in this case, so the solutions should be ``exact'' up to the
machine precision.

\subsubsection{2D Poisson problem}

In the 2D case (Fig.~\ref{fig:2d-sincos}), $u(x, y) \equiv \sin{\left (5 \pi x
  \right )} \cos{\left (5 \pi y \right )}$ was used. In terms of $N_{dof}$,
similar results as in the 1D case were obtained, i.e., increasing the global
continuity of the IGA basis improves the convergence in comparison with
FEM. The same holds for $N_{nz}$ after a certain minimal system resolution.
Also again as in 1D, the standard $C^0$ basis performs better when considering
convergence w.r.t. $N_e$.

\subsubsection{3D Poisson problems}

In 3D, two solutions were considered. The first solution $u(x, y, z) \equiv
\left(x^{3} - 0.125\right) \left(y^{3} - 0.125\right) \left(z^{3} -
  0.125\right)$ (Fig.~\ref{fig:3d-p03}) is a tensor product of order three
polynomials. The results reflect that as a numerical ``zero'' error is obtained
independently of the resolution for the order/degree three bases. The higher
continuity of IGA seems again to mitigate a slight loss of precision for higher
resolutions.

The second solution $u(x, y, z) \equiv \sin{\left (5 \pi x \right )} \sin{\left
    (5 \pi z \right )} \cos{\left (5 \pi y \right )}$
(Fig.~\ref{fig:3d-sincossin}) is a direct generalization of the 2D case, and
behaves in the same way.

The conjugate gradient iterative solver from PETSc \cite{PETSc}, preconditioned
by the incomplete Cholesky decomposition was used in the 2D and 3D cases. The
absolute precision for preconditioned residuals was set to $10^{-18}$, so that
``exact'' solutions are obtained.

\subsection{Simple eigenvalue problems}
\label{sec:evp}

Several simple quantum mechanical systems were considered for our convergence
study, namely an infinite potential well in 2D and 3D, a linear harmonic
oscillator in 2D and 3D, and a hyperbolic 2D potential. The domain was a
unit cube (or square): $\Omega \equiv [-a/2, a/2]^d$, $d = 1, 2, 3$. The
discretization parameters are summarized in Tab.~\ref{tab:evp-pars}. We were
interested in convergence of the two smallest eigenvalues to the analytic
values $\bar \varepsilon_i$. The error was measured using
$\sum_{i=1,2}(|\varepsilon_i - \bar \varepsilon_i|) / 2$.

\begin{table}[ht!]
  \centering
  \begin{tabular}{r|llc}
    system & dimension & $N_e$ & edge length $a$ \\
    \hline
    well & 2D & 20, 50, 100, 250 & 1 \\
    well & 3D & 5, 10, 20, 35 & 1 \\
    oscillator & 2D & 20, 50, 100, 250 & 10 \\
    oscillator & 3D & 5, 10, 20, 35 & 10 \\
    hyperbolic & 2D & 20, 50, 100, 250 & 30
  \end{tabular}
  \caption{Simple eigenvalue problem: discretization parameters --- numbers of
    vertices/knots along the domain edge and the domain edge length.}
  \label{tab:evp-pars}
\end{table}

\begin{figure}[ht!]
  \centering

  \includegraphics[width=0.48\linewidth]{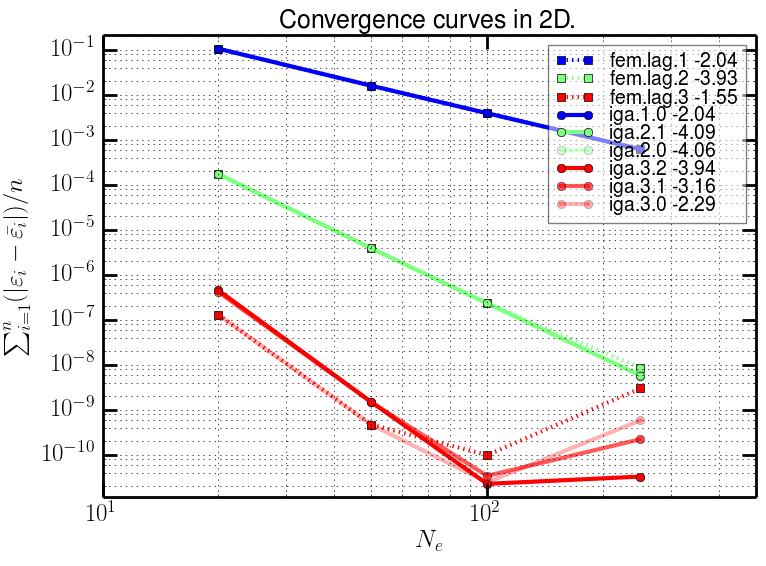}
  \includegraphics[width=0.48\linewidth]{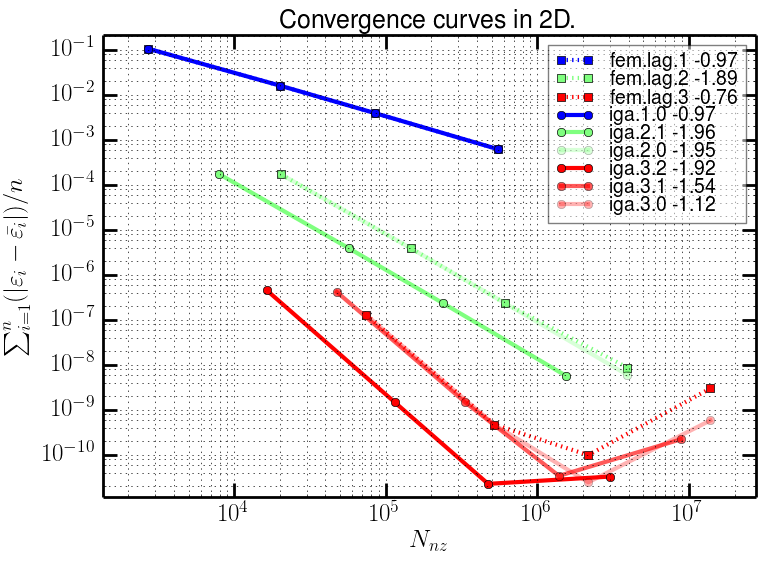}
  \includegraphics[width=0.48\linewidth]{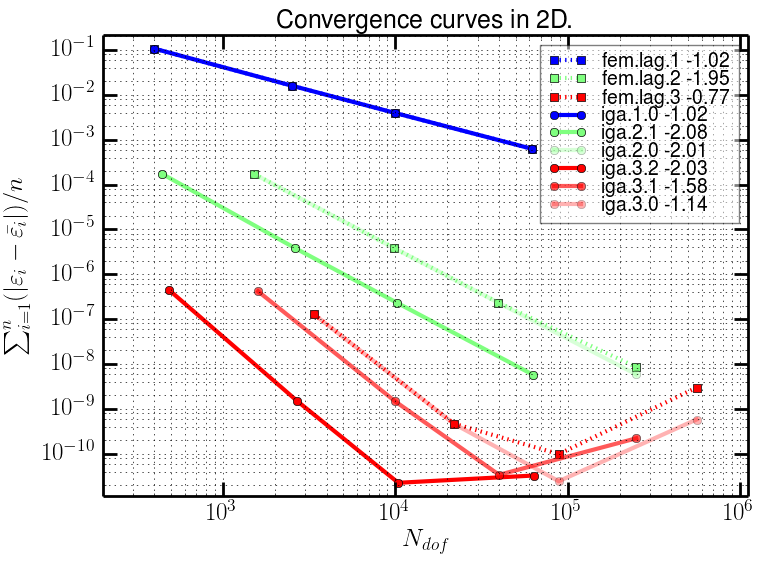}
  \includegraphics[width=0.48\linewidth]{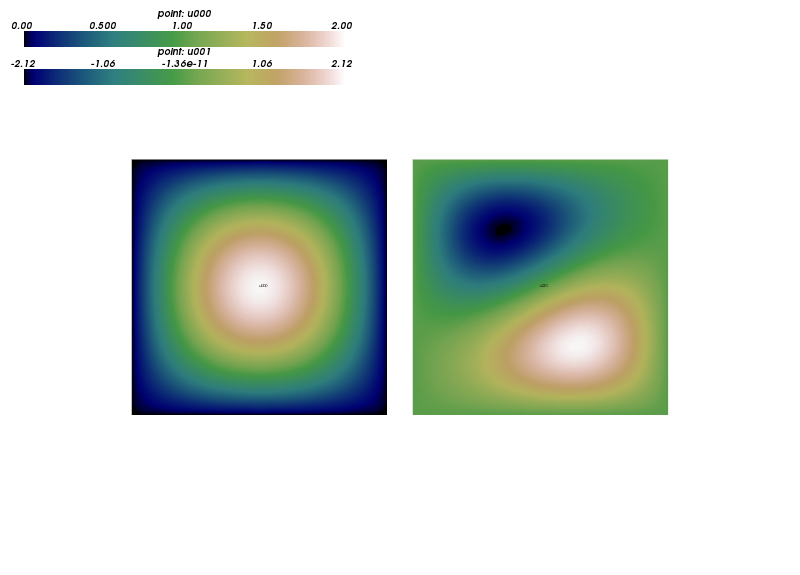}

  \caption{Convergence results for the 2D well
    eigenvalue problem. The
    legend for the FEM shows the polynomial order and the slope of the
    corresponding curve. For the IGA it shows the degree, the global
    continuity, and the slope, respectively. The eigen-functions are shown in
    bottom-right.}
  \label{fig:0-2d-evp-well}
\end{figure}

\begin{figure}[ht!]
  \centering

  \includegraphics[width=0.48\linewidth]{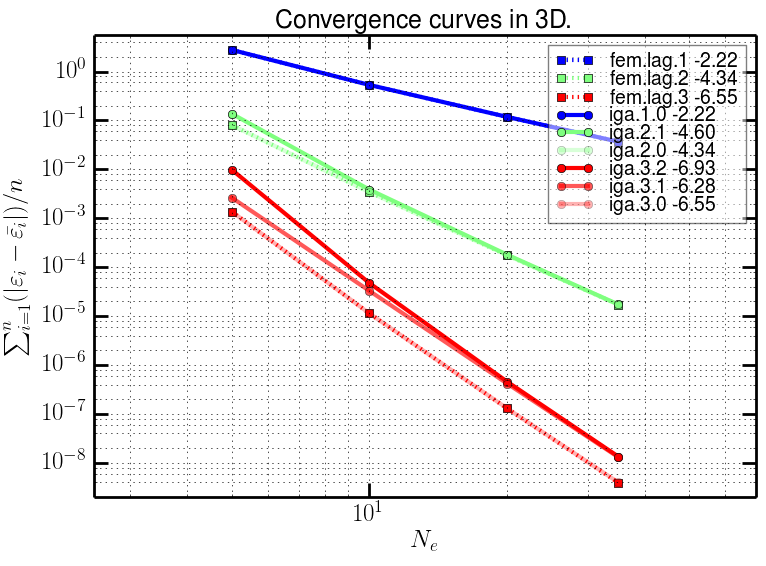}
  \includegraphics[width=0.48\linewidth]{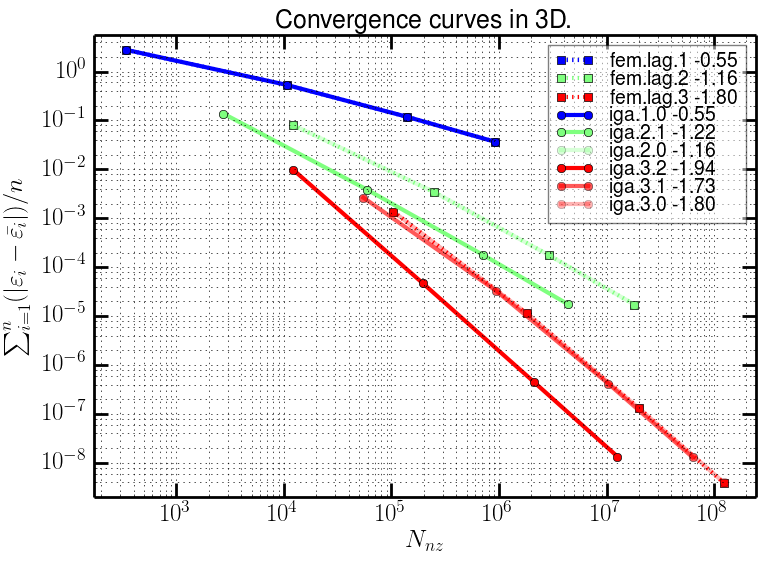}
  \includegraphics[width=0.48\linewidth]{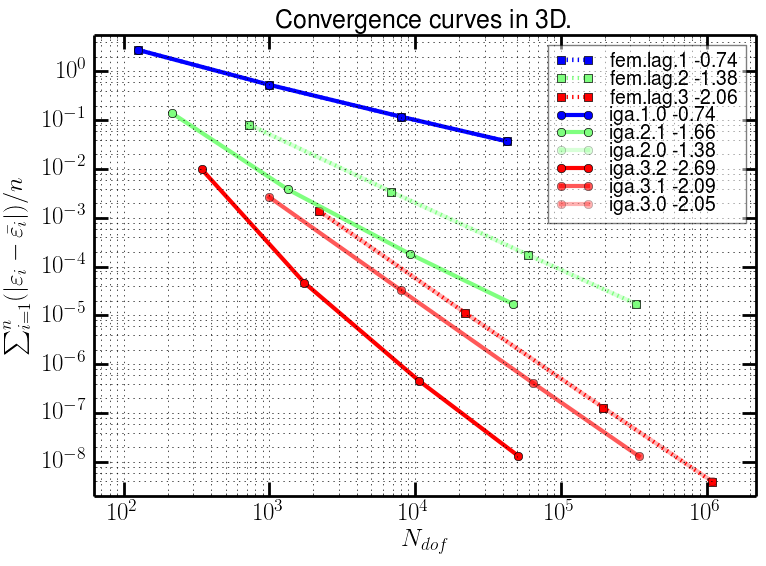}
  \includegraphics[width=0.48\linewidth]{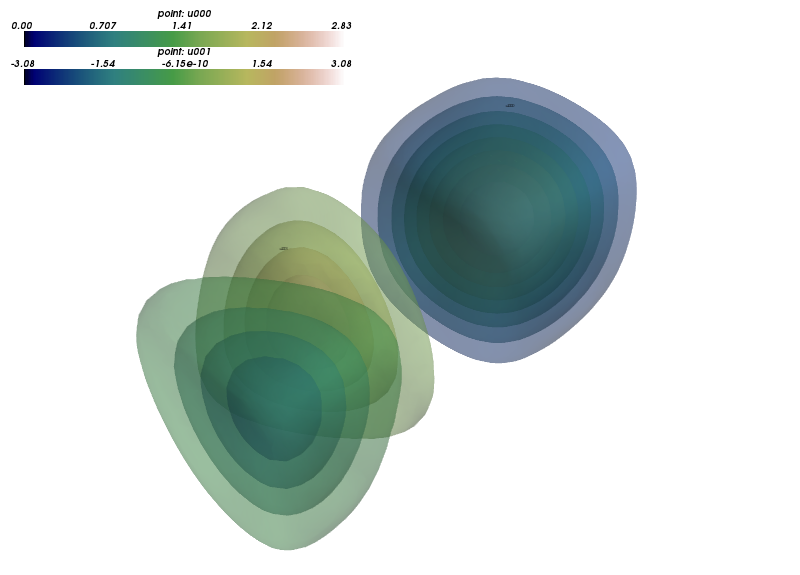}

  \caption{Convergence results for the 3D well
    eigenvalue problem. The
    legend for the FEM shows the polynomial order and the slope of the
    corresponding curve. For the IGA it shows the degree, the global
    continuity, and the slope, respectively. The eigen-functions are shown in
    bottom-right.}
  \label{fig:1-3d-evp-well}
\end{figure}

\begin{figure}[ht!]
  \centering

  \includegraphics[width=0.48\linewidth]{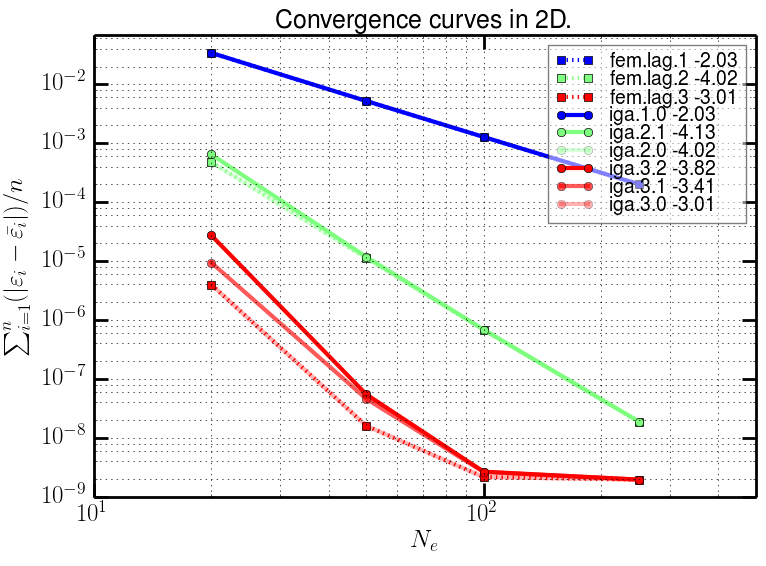}
  \includegraphics[width=0.48\linewidth]{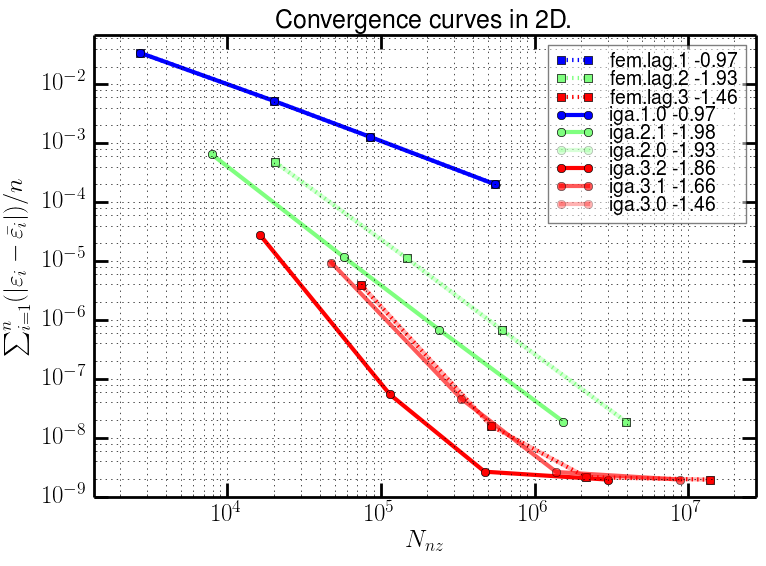}
  \includegraphics[width=0.48\linewidth]{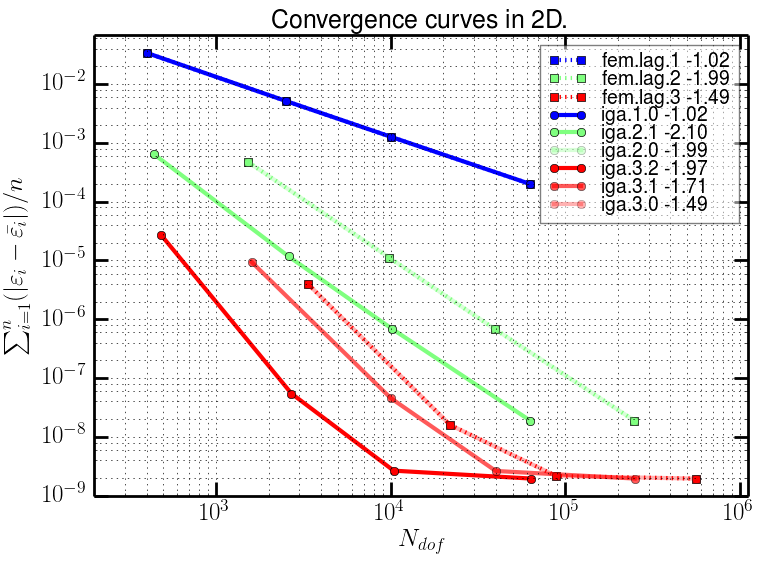}
  \includegraphics[width=0.48\linewidth]{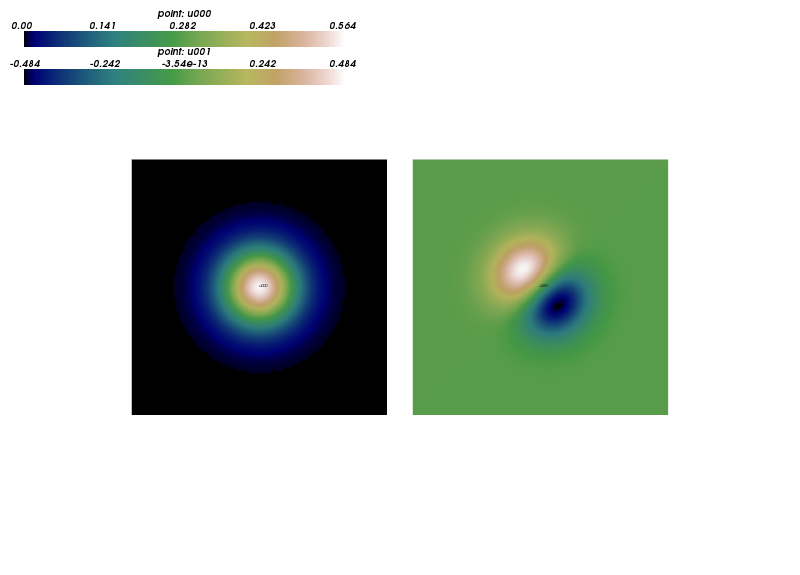}

  \caption{Convergence results for the 2D oscillator
    eigenvalue problem. The
    legend for the FEM shows the polynomial order and the slope of the
    corresponding curve. For the IGA it shows the degree, the global
    continuity, and the slope, respectively. The eigen-functions are shown in
    bottom-right.}
  \label{fig:2-2d-evp-oscillator}
\end{figure}

\begin{figure}[ht!]
  \centering

  \includegraphics[width=0.48\linewidth]{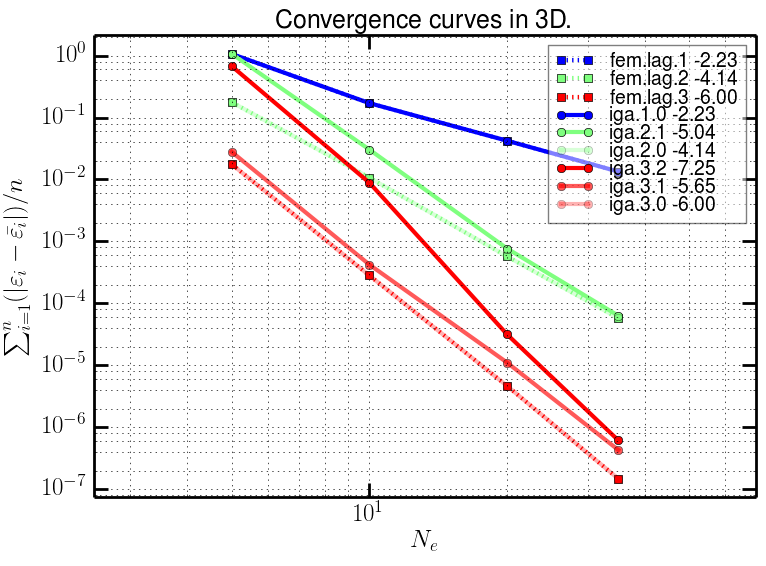}
  \includegraphics[width=0.48\linewidth]{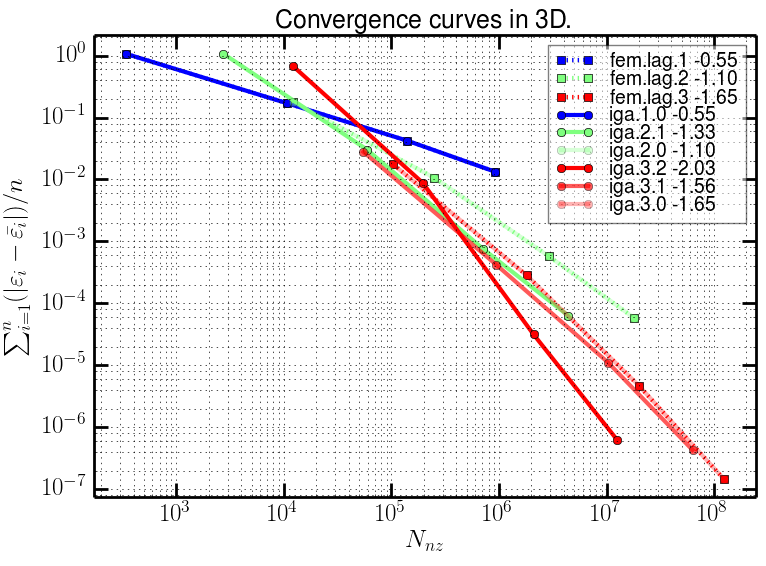}
  \includegraphics[width=0.48\linewidth]{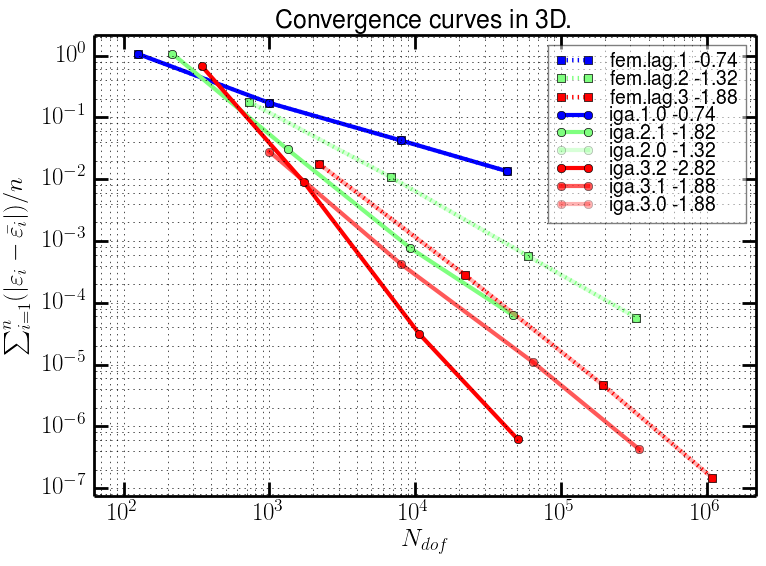}
  \includegraphics[width=0.48\linewidth]{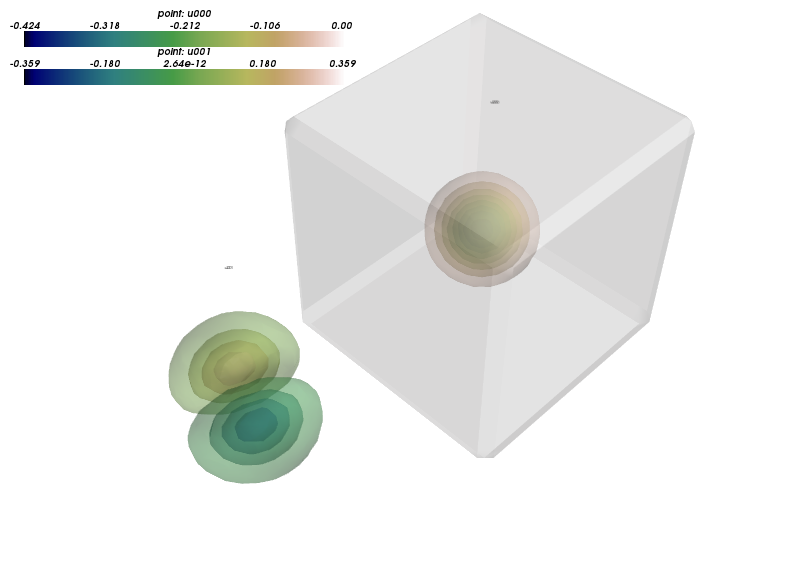}

  \caption{Convergence results for the 3D oscillator
    eigenvalue problem. The
    legend for the FEM shows the polynomial order and the slope of the
    corresponding curve. For the IGA it shows the degree, the global
    continuity, and the slope, respectively. The eigen-functions are shown in
    bottom-right.}
  \label{fig:3-3d-evp-oscillator}
\end{figure}

\begin{figure}[ht!]
  \centering

  \includegraphics[width=0.48\linewidth]{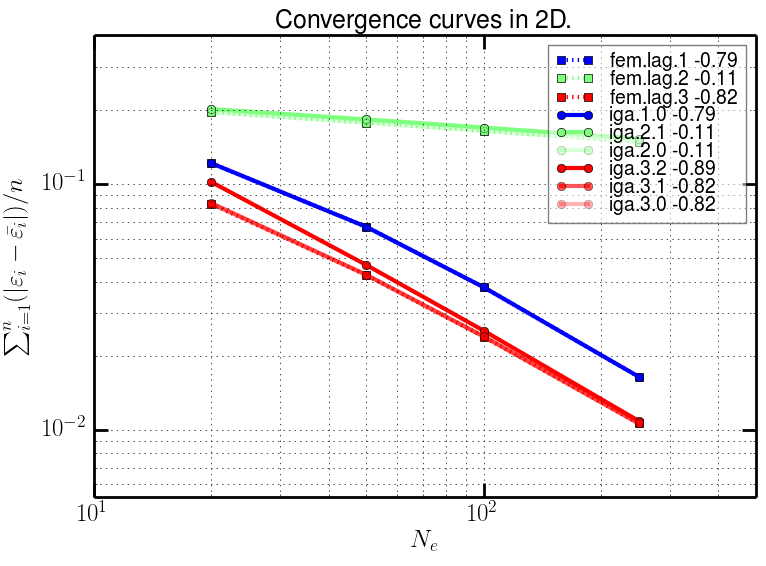}
  \includegraphics[width=0.48\linewidth]{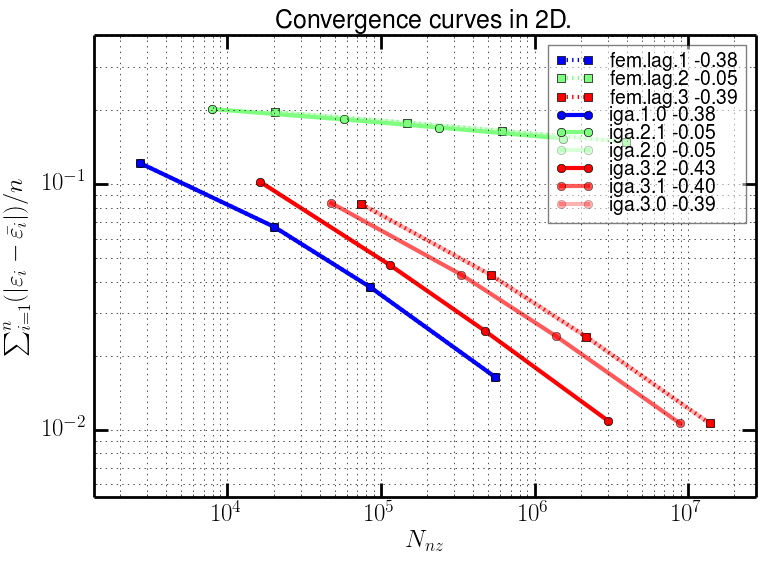}
  \includegraphics[width=0.48\linewidth]{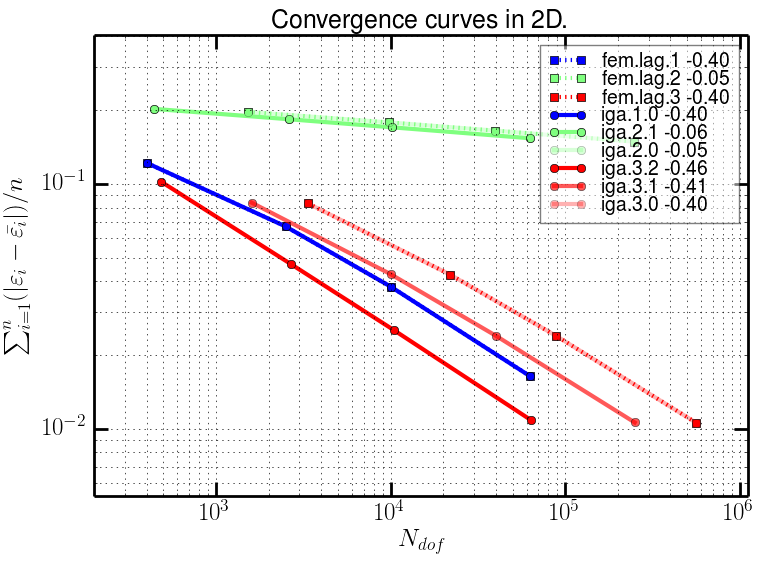}
  \includegraphics[width=0.48\linewidth]{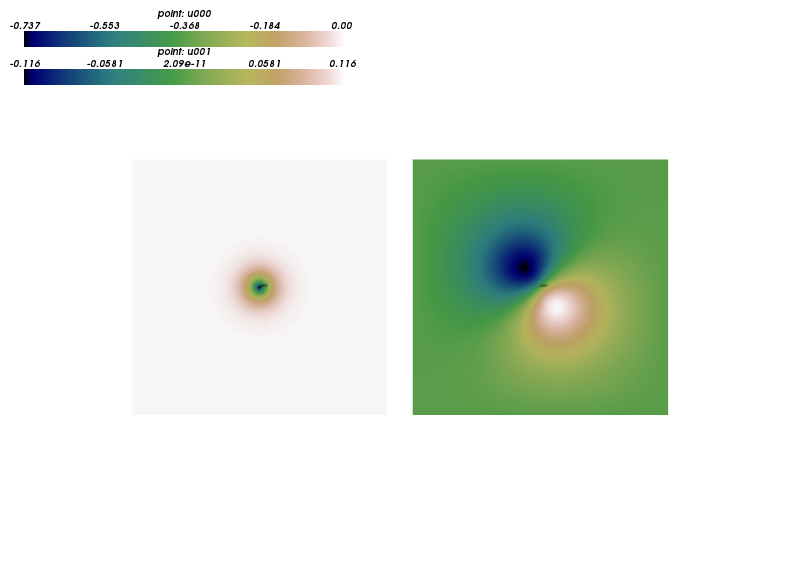}

  \caption{Convergence results for the 2D hyperbolic
    eigenvalue problem. The
    legend for the FEM shows the polynomial order and the slope of the
    corresponding curve. For the IGA it shows the degree, the global
    continuity, and the slope, respectively. The eigen-functions are shown in
    bottom-right.}
  \label{fig:4-2d-evp-hydrogen}
\end{figure}

The convergence of FEM and IGA for the simple eigenvalue problems is compared
in Figs.~\ref{fig:0-2d-evp-well}-~\ref{fig:4-2d-evp-hydrogen}. All the figures
contain the convergence curves w.r.t. $N_e$ (top left), $N_{nz}$ (top right)
and $N_{dof}$ (bottom left). The eigen-functions corresponding to
$\varepsilon_1$, $\varepsilon_2$ are depicted in bottom right. The figure
legends use the same naming scheme as in \ref{sec:poisson}. All eigenvalue
problems in this section were solved using the JDSYM solver from Pysparse
\cite{Pysparse}.

\subsubsection{Infinite potential well}
The potential well can be described by (\ref{eq:kohn-sham-weak}) with a trivial
choice of the potential: $V(\rb) \equiv 0$. The exact eigenvalues are given by
\begin{eqnarray*}
  \bar \varepsilon_i = \frac{\pi^2}{2 a^2} c_i \mbox{ , where }
  c_i = \smatrix{c}{\{} 2, 5 \mbox{ in 2D,} \\ 3, 6 \mbox{ in 3D.} \ematrix{.}
\end{eqnarray*}
In our case, $\bar \varepsilon_i$ was equal to 9.869604401089358,
24.674011002723397 in 2D and to 14.804406601634037, 29.608813203268074 in 3D.

The convergence curves for the 2D case are shown in
Fig.~\ref{fig:0-2d-evp-well}.  Here, IGA with increasing continuity performs
progressively better then FEM with the same polynomial order when the
convergence w.r.t. $N_{nz}$ and $N_{dof}$ is considered. On the other hand, its
error is slightly higher than the FEM or $C^0$ IGA error for the $N_e$
curve. The right parts of the curves for order/degree three solutions exhibit a
loss numerical precision for very small errors --- this loss of precision seems
to decrease with increasing the IGA basis continuity, and for the highest
system resolution, IGA outperforms FEM even w.r.t. $N_e$. This is probably
related to a better conditioning of the resulting eigenvalue problem. Compare
to the 1D Poisson problem example in Section~\ref{sec:poisson}, where also the
convergence to machine precision limits was achieved.

The convergence curves for the 3D case are shown in
Fig.~\ref{fig:1-3d-evp-well}.  The results are qualitatively analogous to the
2D case, but without the loss of numerical precision, because the machine
precision limit was not reached.

\subsubsection{Linear harmonic oscillator}
The linear harmonic oscillator can be described by (\ref{eq:kohn-sham-weak})
with a particular choice of the potential: $V(\rb) \equiv \frac{1}{2} \rb^2$.
The exact eigenvalues $\bar \varepsilon_i$ are equal to 1, 2 in 2D and to 1.5,
2.5 in 3D.

The convergence curves for the 2D case are shown in
Fig.~\ref{fig:2-2d-evp-oscillator} and are analogous to the 2D potential well
results, with the exception that the higher continuity of IGA does not help to
fight the machine precision limit.

The convergence curves for the 3D case are shown in
Fig.~\ref{fig:3-3d-evp-oscillator} and qualitatively correspond to the 3D
potential well results.

\subsubsection{Hyperbolic 2D potential}

The hyperbolic 2D potential serves as a non-physical 2D analogy of Coulombic
potential for hydrogen atom. It can be described by (\ref{eq:kohn-sham-weak})
with a particular choice of the potential: $V(\rb) \equiv -\frac{1}{2\rb}$.
The exact eigenvalues $\bar \varepsilon_i$ are equal to -0.5, $-0.0\bar{5}$,
according to
\begin{displaymath}
  \bar \varepsilon_i = -\frac{1}{8(i-0.5)^2} \mbox{ , where }
  i = 1, 2 \;.
\end{displaymath}
Unlike the previous examples, the potential $V(\rb)$ has a singularity at $\rb
= 0$. To avoid numerical problems, the singularity removed by setting the
radii smaller than $10^{-6}$ to $10^{-6}$. Due to that, the numerical solutions
converge to slightly different values than given above. Nevertheless, as can be
seen in Fig.~\ref{fig:4-2d-evp-hydrogen}, the convergence of IGA solution with
high global continuity seems to be better than that of FEM in terms of $N_e$
and $N_{dof}$, and worse in case of $N_{nz}$.

Note that the singularity problem is not present in full DFT scheme below,
thanks to the use of carefully chosen pseudopotentials.

\subsection{DFT loop}
\label{sec:dft}

The computations were done on a cube domain having the edge size of 14 atomic
units, with varying number of vertices/knots along an edge. A nitrogen atom was
used for the benchmark due to availability of reference solution values for
this simple system.

\begin{figure}[ht!]
  \centering

  \includegraphics[width=0.48\linewidth]
  {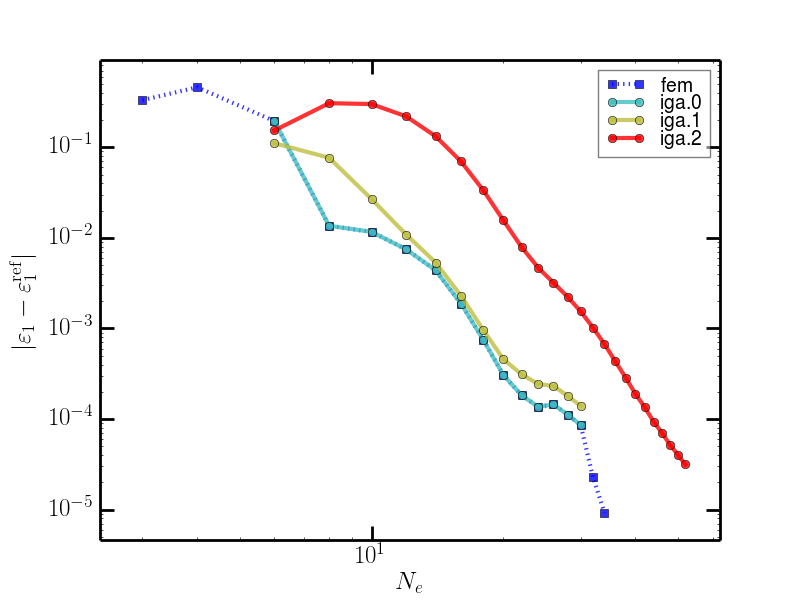}
  \includegraphics[width=0.48\linewidth]
  {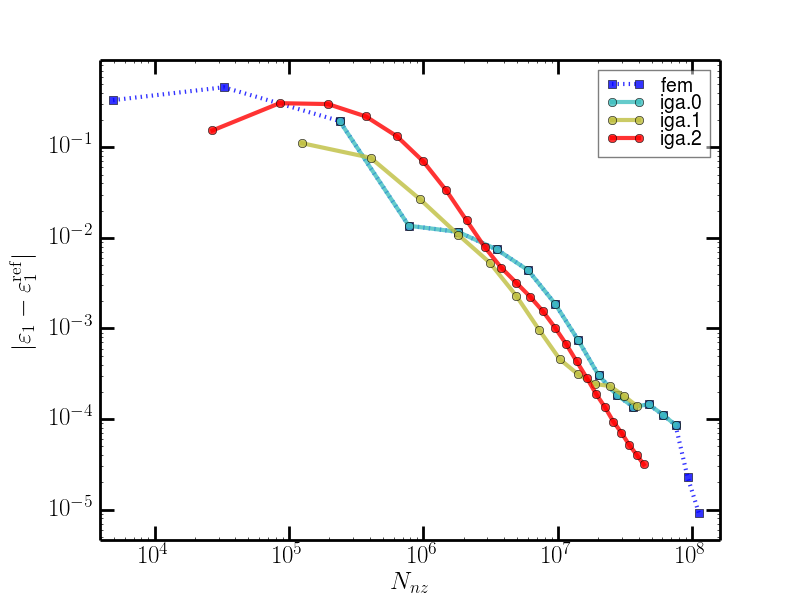}

  \includegraphics[width=0.48\linewidth]
  {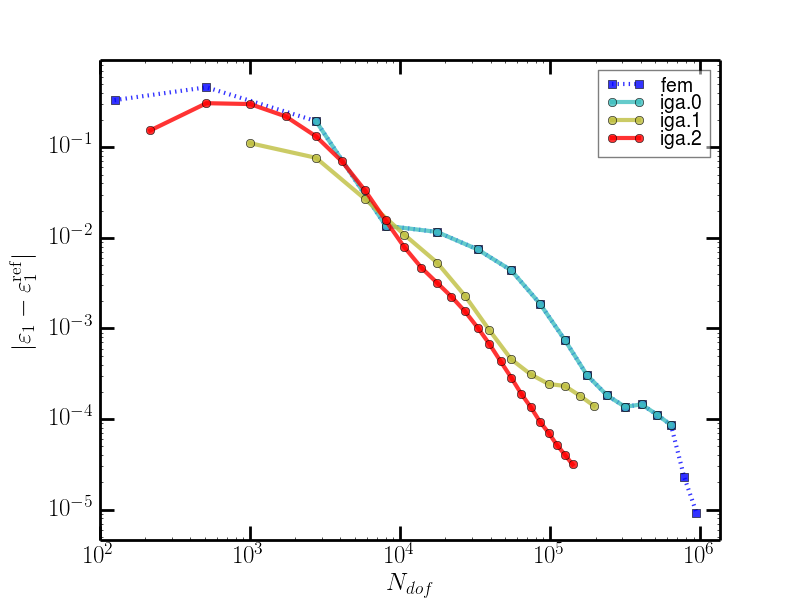}

  \caption{Convergence of the first eigenvalue $\varepsilon_1$: difference
    w.r.t. a reference value. The IGA labels indicate the global basis
    continuity.}
  \label{fig:conv-dft-eig1}
\end{figure}

\begin{figure}[ht!]
  \centering

  \includegraphics[width=0.48\linewidth]
  {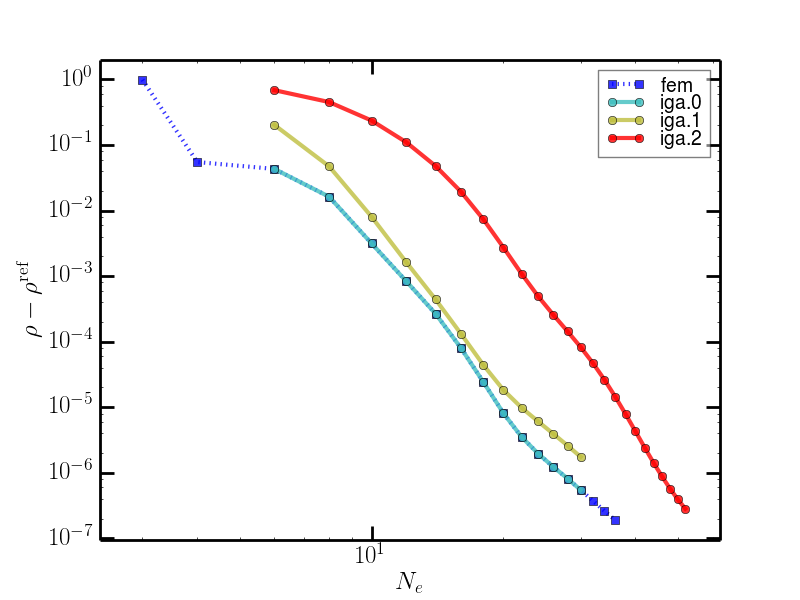}
  \includegraphics[width=0.48\linewidth]
  {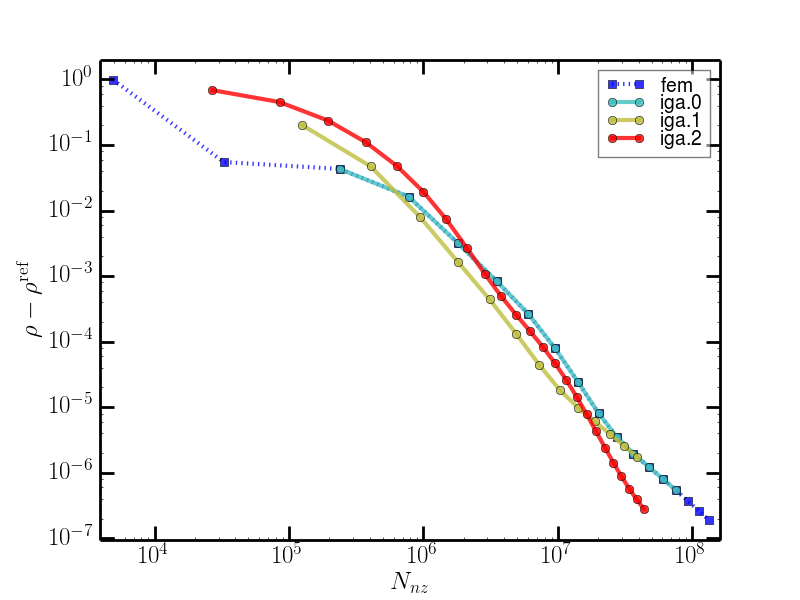}

  \includegraphics[width=0.48\linewidth]
  {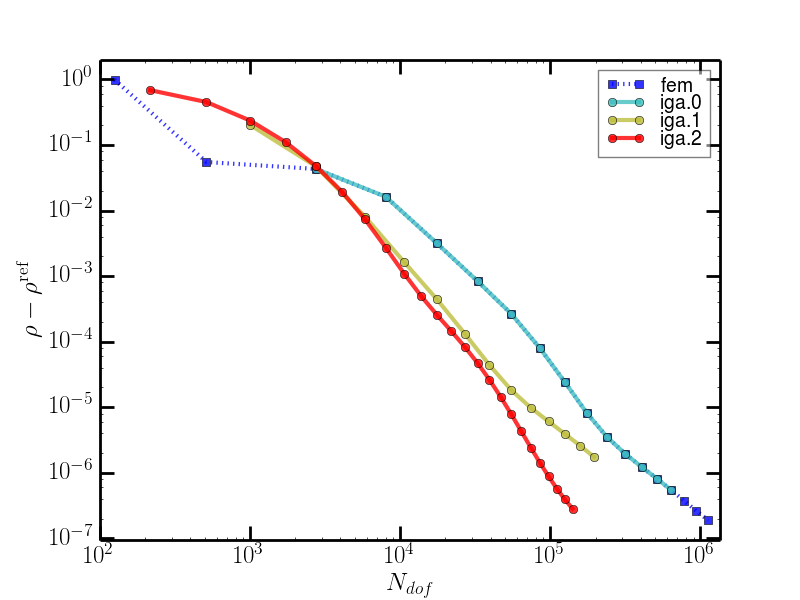}

  \caption{Convergence of the charge density $\rho$: difference w.r.t. a
    reference value. The IGA labels indicate the global basis continuity.}
  \label{fig:conv-dft-rho}
\end{figure}

The tri-cubic FEM basis and degree three IGA basis were used. The continuities
$C^0$, $C^1$ and $C^2$ were used for the IGA basis. In
Fig.~\ref{fig:conv-dft-eig1} we can see the convergence of the first eigenvalue
$\varepsilon_1$ of problem (\ref{eq:kohn-sham-weak}) to a reference value. The
Fig.~\ref{fig:conv-dft-rho} depicts the convergence of the charge density
$\rho$ to a reference value.

The following observations can be made from both figures. As expected, the IGA
with $C^0$ continuity is exactly equivalent to the FEM. The higher degree of
continuity in IGA leads to an improved convergence w.r.t. $N_{nz}$ and
$N_{dof}$, and a worse convergence w.r.t. $N_e$.

\subsection{Results Summary}
\label{sec:summary}

We were interested in convergence w.r.t. three parameters: the number of
vertices/knots along the domain edge $N_e$ ($\approx$ cost of quadrature and
assembling), the number of non-zero entries in the matrices $N_{nz}$ ($\approx$
cost of matrix-vector products) and the total number of degrees of freedom
$N_{dof}$.  The convergence of the Poisson equation solution to a manufactured
analytical solution, the convergence of two smallest eigenvalues of simple
quantum mechanical systems and finally the convergence of the complete DFT loop
were assessed.

As reported in Section~\ref{sec:poisson}, in all tests, IGA with $C^2$ global
continuity is the most efficient in term of convergence w.r.t. $N_{dof}$, i.e.,
the sizes of the vectors and matrices involved, which relate to the difficulty
of solving the Poisson equation (\ref{eq:poisson-weak}) or the eigenvalue
problem (\ref{eq:kohn-sham-weak}).

In the Poisson equation test problems, IGA with increasing continuity performs
progressively better then FEM with the same polynomial order when the
convergence w.r.t. $N_{nz}$ and $N_{dof}$ is considered. On the other hand, the
standard $C^0$ basis performs better when considering convergence w.r.t. $N_e$.

Similar results were obtained in eigenvalue problems in Section~\ref{sec:evp}
originating from simple quantum mechanical systems. Again, IGA with increasing
continuity performs progressively better then FEM with the same polynomial
order considering convergence w.r.t. $N_{nz}$ and $N_{dof}$, and IGA error is
slightly higher than FEM or $C^0$ IGA errors for the $N_e$ curves.

Considering the convergence of the eigenvalue $\varepsilon_1$ and the charge
density $\rho$ computed by the complete DFT loop in Section~\ref{sec:dft},
increasing the IGA basis continuity improves again the convergence
w.r.t. $N_{nz}$ and $N_{dof}$.  In the case of the convergence w.r.t. $N_e$,
however, increasing the basis continuity (and thus decreasing $N_{dof}$ for the
same number of elements) leads to a significantly worse convergence. This
corresponds to a much higher numerical quadrature cost of $C^2$ IGA basis when
compared to a $C^0$ basis of the same accuracy.

\section{Conclusion}

We compared numerical convergence properties of FEM and IGA using problems
originating from various stages of our electronic structure calculation
algorithm, based on the density functional theory, the environment-reflecting
pseudopotentials and a weak solution of the Kohn-Sham equations. Our computer
implementation built upon the open source package SfePy supports computations
both with the FE basis and the NURBS or B-splines basis of IGA. The latter
allows a high global continuity in approximation of unknown fields, so
convergence properties of B-spline bases with global continuities up to $C^2$
were examined, because having a globally $C^2$ continuous approximation is
crucial for efficient computing of derivatives of the total energy
w.r.t. atomic positions etc., as given by the Hellmann-Feynman
theorem~\cite{IGA-FEM-Cimrman-1}.

Overall, the results summarized in Section~\ref{sec:summary} support our choice
of IGA as a viable alternative to FEM in electronic structure calculations. To
alleviate the numerical quadrature cost, reduced quadrature rules has been
proposed for the context of the Bézier extraction \cite{IGA-Schillinger-2},
which we plan to assess in future.

\noindent
\textbf{Acknowledgments:} The work was supported by the Grant Agency of the
Czech Republic, project P108/11/0853. R. Kolman's work was supported by the
grant project of the Czech Science Foundation (GACR), No. GAP 101/12/2315,
within the institutional support RVO:61388998.

\section*{References}

\bibliography{convergence2015-iga-dft.bib}

\end{document}